\documentclass[12pt, reqno]{amsart} 
\usepackage{amssymb,latexsym,amsmath,enumerate,nicefrac}
\usepackage{amsaddr}
\usepackage{float}

\usepackage{caption}

\usepackage{color}
\definecolor{myurlcolor}{rgb}{0.6,0,0}
\definecolor{mycitecolor}{rgb}{0,0,0.8}
\definecolor{myrefcolor}{rgb}{0,0,0.8}
\usepackage[bookmarks=false,pagebackref=true]{hyperref}
\hypersetup{colorlinks,
linkcolor=myrefcolor,
citecolor=mycitecolor,
urlcolor=myurlcolor}

\newtheorem{theorem}{Theorem}[section]
\newtheorem{lemma}[theorem]{Lemma}

\newtheorem{proposition}[theorem]{Proposition}
\newtheorem*{corollary}{Corollary}
\theoremstyle{definition}
\newtheorem{definition}[theorem]{Definition}

\theoremstyle{remark}

\usepackage{comment}
\usepackage{tikz}

\usetikzlibrary{arrows}
 \usetikzlibrary{matrix,positioning}

\allowdisplaybreaks
\includecomment{pf}

\excludecomment{expo}

\excludecomment{cut}

\makeatletter
\newcommand*\bigcdot{\mathpalette\bigcdot@{.6}}
\newcommand*\bigcdot@[2]{\mathbin{\vcenter{\hbox{\scalebox{#2}{$\m@th#1\bullet$}}}}}
\makeatother

\newcounter{BWYtable}

\newcounter{BWYDiagram}

\newcounter{BWYFigure}

\newcommand{\1}{1}
\newcommand{\2}{2}
\newcommand{\3}{3}
\newcommand{\4}{4}

\newcommand{\C}{\mathscr{C}}

\usepackage{tikz}
\usetikzlibrary{calc}
\usepackage{graphics}
\usepackage{dsfont}
\usepackage{amsmath}
\usepackage{amsfonts,amssymb}
\usepackage{mathrsfs}
\usepackage{mathtools}
\usepackage{comment}

\renewcommand{\(}{\left(} 
\renewcommand{\)}{\right)}
\newcommand{\lpar}{{\textnormal(\hspace{-.5pt}}}
\newcommand{\rpar}{{\textnormal)\,}}

\newcommand{\catname}[1]{{\text{\sffamily #1}}}

\newcommand{\Set}{\catname{Set}}

\newcommand{\Dif}{{\catname{Diff}}}
\newcommand{\SurjSub}{{\catname{SurjSub}}}
\newcommand{\Riem}{{\catname{RiemSurj}}}
\newcommand{\SympSurj}{{\catname{SympSurj}}}
\newcommand{\Cat}{{\mathfrak M}}
\newcommand{\HamSy}{{\catname{HamSy}}}
\newcommand{\LagSy}{{\catname{LagSy}}}
\newcommand{\Top}{{\catname{Top}}}

\newcommand{\F}{\mathcal F}

\newcommand{\SA}{S_A}
\newcommand{\SL}{S_L}
\newcommand{\SR}{S_R}
\newcommand{\sL}{s_L}
\newcommand{\sR}{s_R}

\newcommand{\TA}{T_A}
\newcommand{\TL}{T_L}
\newcommand{\TR}{T_R}
\newcommand{\tL}{t_L}
\newcommand{\tR}{t_R}

\newcommand{\CA}{C_A}
\newcommand{\CL}{C_L}
\newcommand{\CR}{C_R}
\newcommand{\cL}{c_L}
\newcommand{\cR}{c_R}

\newcommand{\QA}{Q_A}
\newcommand{\QL}{Q_L}
\newcommand{\QR}{Q_R}
\newcommand{\qL}{q_L}
\newcommand{\qR}{q_R}
\newcommand{\qM}{q_M}

\newcommand{\PA}{P_A}
\newcommand{\PL}{P_L}
\newcommand{\PR}{P_R}
\newcommand{\pL}{p_L}
\newcommand{\pR}{p_R}

\renewcommand{\d}{{\rm d}}

\newcommand{\Lag}{{\mathcal L}}

\newcommand{\FLag}{\mathscr{L}}
\newcommand{\preFLag}{\mathscr{K}}
\newcommand{\g}{\flat}
\newcommand{\ginv}{\sharp}
\newcommand{\dualmet}[1]{g^{\ast}_{#1}}
\newcommand{\met}[1]{g_{#1}}
\newcommand{\proj}[1]{\rho_{#1}}
\newcommand{\dualproj}[1]{\pi_{#1}}
\newcommand{\tform}[1]{\omega_{#1}}

\def\mathrlap{\mathpalette\mathrlapinternal}

\def\mathrlapinternal#1#2{\rlap{$\mathsurround=0pt#1{#2}$}}

\usetikzlibrary{decorations.pathmorphing}
\usetikzlibrary{arrows}
\usetikzlibrary{calc}
\usepackage{graphics}
\usepackage{dsfont}
\usepackage{amsmath}
\usepackage{amsfonts,amssymb}
\usepackage{mathrsfs}
\usepackage{mathtools}

\pgfkeys{/tikz/.cd, arrowhead/.default=-1pt, arrowhead/.code={}}

\includecomment{showTikz}

\excludecomment{showPDF}

\excludecomment{showTOC}

\excludecomment{showData}

\begin{document}

\title{Open Systems in Classical Mechanics}



\author{John C. Baez$^1$, David Weisbart$^2$, \and Adam M. Yassine$^3$}
\address{$^{1,2}$Department of Mathematics\\University of California, Riverside\\Riverside, California\\USA}\address{$^3$Department of Mathematics and Actuarial Science\\The American University in Cairo\\New Cairo, Egypt}\email{$^1$baez@math.ucr.edu}\email{$^2$weisbart@math.ucr.edu} \email{$^3$adam.yassine@aucegypt.edu}

\maketitle

\pagestyle{plain}

\begin{abstract} 
Generalized span categories provide a framework for formalizing mathematical models of open systems in classical mechanics. We introduce categories $\LagSy$ and $\HamSy$ that respectively provide a categorical framework for the Lagrangian and Hamiltonian descriptions of open classical mechanical systems.  The morphisms of $\LagSy$ and $\HamSy$ correspond to such open systems, and composition of morphisms models the construction of systems from subsystems.  The Legendre transformation gives rise to a functor from $\LagSy$ to $\HamSy$ that translates from the Lagrangian to the Hamiltonian perspective.
\end{abstract}



\begin{showTOC}
\tableofcontents
\end{showTOC}

\section{Introduction}\label{sec:intro}

Category theory provides a formalism for unifying ideas across a wide spectrum of disciplines.  The last few decades have seen the emergence of applied category theory \cite{FongSpivak,Spiv}. One prominent program in this subject is to describe ``open'' systems---that is, systems that can interact with their surroundings---as morphisms in appropriate categories, where composition describes how open systems can be combined to form larger systems.

The idea of describing open systems as morphisms arose from extended topological quantum field theory, where the manifold describing space can be built up by composing cobordisms, manifolds  with boundary that describe smaller regions of space \cite{Baez-Dolan,BFV,Freed,Hau}.   It was later applied in a more down-to-earth way to electrical circuits \cite{BCR,BF}, Markov processes \cite{BFP}, and a wide variety of dynamical systems \cite{BP,LS,SSV}.   The morphisms in these categories are often spans or cospans with extra structure, and there are now several formalisms for constructing such categories \cite{CourserThesis}.

Our goal here is to apply this idea to Lagrangian and Hamiltonian mechanics, and describe the Legendre transformation as as a functor from a category with open Lagrangian systems as morphisms to a similar category of open Hamiltonian systems.  Since the study of classical systems involves solving differential equations that describe paths on general Riemannian and symplectic manifolds, it is in some ways more complicated than the examples treated earlier.  The current work investigates some previously unidentified structures that appear critical to the study of open systems in classical mechanics.

The systems under consideration have a state space that is either the tangent bundle to a Riemannian manifold in the Lagrangian description or a symplectic manifold in the Hamiltonian description \cite{Arn}.  A path in the state space models the motion of the system.   The state space of any subsystem is a quotient space of that of the entire system.   For Lagrangian systems we require that the quotient maps be surjective Riemannian submersions.  For Hamiltonian systems, we require that they be surjective Poisson maps between symplectic manifolds.

A simple example is this system with three point masses attached by springs, where all motion is along the same line:
\begin{center}
\begin{showTikz}
\begin{tikzpicture}
%
\coordinate(r1) at (-1.5,0);
\coordinate(r2) at (0,0);
\coordinate(r3) at (1.5,0);
{\draw[decorate,decoration={coil,segment length=4pt, amplitude = 4pt},rotate=0] (r1)  -- (r2);}
{\draw[decorate,decoration={coil,segment length=4pt, amplitude = 4pt},rotate=0] (r2)  -- (r3);}
{\draw[fill= black] (r1) circle (2.5pt);}
{\draw[fill= black] (r2) circle (2.5pt);}
{\draw[fill= black] (r3) circle (2.5pt);}
\end{tikzpicture}
\end{showTikz}
\begin{showPDF}
\includegraphics{Figure1.pdf}
\end{showPDF}
\end{center}
We can build a complicated system by attaching additional point masses and springs in series:
\begin{center}
\begin{showTikz}
\begin{tikzpicture}

%
\coordinate(r1) at (-1.5,0);
\coordinate(r2) at (0,0);
\coordinate(r3) at (1.5,0);
\coordinate(r4) at (3,0);
\coordinate(r5) at (4.5,0);

\coordinate(dotsa) at (2,0);
\coordinate(dotsb) at (2.5,0);

\coordinate(dots1) at (2.15,0);
\coordinate(dots2) at (2.25,0);
\coordinate(dots3) at (2.35,0);

{\draw[decorate,decoration={coil,segment length=4pt, amplitude = 4pt},rotate=0] (r1)  -- (r2);}
{\draw[decorate,decoration={coil,segment length=4pt, amplitude = 4pt},rotate=0] (r2)  -- (r3);}

{\draw[decorate,decoration={coil,segment length=4pt, amplitude = 4pt},rotate=0] (r3)  -- (dotsa);}
{\draw[decorate,decoration={coil,segment length=4pt, amplitude = 4pt},rotate=0] (dotsb)  -- (r4);}

{\draw[decorate,decoration={coil,segment length=4pt, amplitude = 4pt},rotate=0] (r4)  -- (r5);}

{\draw[fill= black] (r1) circle (2.5pt);}
{\draw[fill= black] (r2) circle (2.5pt);}
{\draw[fill= black] (r3) circle (2.5pt);}
{\draw[fill= black] (r4) circle (2.5pt);}
{\draw[fill= black] (r5) circle (2.5pt);}

{\draw[fill= black] (dots1) circle (.5pt);}
{\draw[fill= black] (dots2) circle (.5pt);}
{\draw[fill= black] (dots3) circle (.5pt);}
\end{tikzpicture}
\end{showTikz}
\begin{showPDF}
\includegraphics{Figure2.pdf}
\end{showPDF}
\end{center}
View a pair of point masses attached by a spring as a fundamental component, or subsystem, of one of these more complicated systems.  These subsystems are then ``open systems'', in the sense that both forces internal to the subsystem and external forces of the larger system govern the dynamics of the subsystems. 

We may build such a larger system out of two subsystems by identifying the right mass of the subsystem on the left with the left mass of the subsystem on the right, as follows:
\begin{center}
\begin{showTikz}
\begin{tikzpicture}


\coordinate(r1) at (-1.5,3);
\coordinate(r2) at (0,3);
\coordinate(r3) at (1.5,3);

\coordinate(r4) at (-1.5,1.5);
\coordinate(ra) at (-2.25,1.5);
\coordinate(rb) at (-.75,1.5);

\coordinate(r5) at (1.5,1.5);
\coordinate(rc) at (.75,1.5);
\coordinate(rd) at (2.25,1.5);

\coordinate(br) at (-3,0);
\coordinate(bm) at (0,0);
\coordinate(bl) at (3,0);

\coordinate(l1) at (-.75,3);
\coordinate(l2) at (.75,3);
\coordinate(l3) at (-1.5,1.5);
\coordinate(l4) at (1.5,1.5);
{\draw[decorate,decoration={coil,segment length=4pt, amplitude = 4pt},rotate=0] (r1)  -- (r2);}
{\draw[decorate,decoration={coil,segment length=4pt, amplitude = 4pt},rotate=0] (r2)  -- (r3);}

{\draw[decorate,decoration={coil,segment length=4pt, amplitude = 4pt},rotate=0] (ra)  -- (rb);}
{\draw[decorate,decoration={coil,segment length=4pt, amplitude = 4pt},rotate=0] (rc)  -- (rd);}

{\draw[fill= black] (r1) circle (2.5pt);}
{\draw[fill= black] (r2) circle (2.5pt);}
{\draw[fill= black] (r3) circle (2.5pt);}

\draw[fill= black] (ra) circle (2.5pt);
\draw[fill= black] (rb) circle (2.5pt);
\draw[fill= black] (rc) circle (2.5pt);
\draw[fill= black] (rd) circle (2.5pt);

\draw[fill= black] (br) circle (2.5pt);
\draw[fill= black] (bm) circle (2.5pt);
\draw[fill= black] (bl) circle (2.5pt);

{\draw[shorten >=18pt,shorten <=15pt, ->, arrowhead=5pt, line width=.5pt] (r2) -- (r4);}
{\draw[shorten >=18pt,shorten <=15pt, ->, arrowhead=5pt, line width=.5pt] (r2) -- (r5);}
{\draw[shorten >=10pt,shorten <=15pt, ->, arrowhead=5pt, line width=.5pt] (r4) -- (br);}
{\draw[shorten >=10pt,shorten <=15pt, ->, arrowhead=5pt, line width=.5pt] (r4) -- (bm);}
{\draw[shorten >=10pt,shorten <=15pt, ->, arrowhead=5pt, line width=.5pt] (r5) -- (bm);}
{\draw[shorten >=10pt,shorten <=15pt, ->, arrowhead=5pt, line width=.5pt] (r5) -- (bl);}

\end{tikzpicture}
\end{showTikz}
\begin{showPDF}
\includegraphics{Figure3.pdf}
\end{showPDF}
\end{center}
Here we depict the state spaces of these systems from a Hamiltonian perspective: 
\begin{center}
\begin{showTikz}
\begin{tikzpicture}

\coordinate(r1) at (-1.5,3);
\coordinate(r2) at (0,3);
\coordinate(r3) at (1.5,3);

\coordinate(r4) at (-1.5,1.5);
\coordinate(ra) at (-2.25,1.5);
\coordinate(rb) at (-.75,1.5);

\coordinate(r5) at (1.5,1.5);
\coordinate(rc) at (.75,1.5);
\coordinate(rd) at (2.25,1.5);

\coordinate(br) at (-3,0);
\coordinate(bm) at (0,0);
\coordinate(bl) at (3,0);

\coordinate(l1) at (-.75,3);
\coordinate(l2) at (.75,3);
\coordinate(l3) at (-1.5,1.5);
\coordinate(l4) at (1.5,1.5);

\node[] at (r2) {$T^\ast {\mathbb R^2}\times_{T^\ast{\mathbb R}} T^\ast {\mathbb R^2}$};
\node[] at (l3) {$T^\ast {\mathbb R^2}$};
\node[] at (l4) {$T^\ast {\mathbb R^2}$};
\node[] at (bl) {$T^\ast {\mathbb R}$};
\node[] at (bm) {$T^\ast {\mathbb R}$};
\node[] at (br) {$T^\ast {\mathbb R}$};

{\draw[shorten >=18pt,shorten <=15pt, ->, arrowhead=5pt, line width=.5pt] (r2) -- (r4);}
{\draw[shorten >=18pt,shorten <=15pt, ->, arrowhead=5pt, line width=.5pt] (r2) -- (r5);}
{\draw[shorten >=10pt,shorten <=15pt, ->, arrowhead=5pt, line width=.5pt] (r4) -- (br);}
{\draw[shorten >=10pt,shorten <=15pt, ->, arrowhead=5pt, line width=.5pt] (r4) -- (bm);}
{\draw[shorten >=10pt,shorten <=15pt, ->, arrowhead=5pt, line width=.5pt] (r5) -- (bm);}
{\draw[shorten >=10pt,shorten <=15pt, ->, arrowhead=5pt, line width=.5pt] (r5) -- (bl);}

\end{tikzpicture}
\end{showTikz}
\begin{showPDF}
\includegraphics{Figure4.pdf}
\end{showPDF}
\end{center}
Each of the maps  is a canonical projection, and a surjective Poisson map between symplectic manifolds.   At the lowest level are the state spaces of the three distinct masses.  If each mass moves along a line then each system has $T^\ast\mathbb R$, the cotangent bundle to $\mathbb R$, as its state space.  At the middle level are two spring-mass systems, each with a state space given by $T^\ast\mathbb R^2$.  On the top level, the total system consists of three masses interacting in series.  The state space for this total system is a fibered product of two copies of the symplectic manifold $T^\ast\mathbb R^2$ over the manifold $T^\ast\mathbb R$.  

The fibered product is a six dimensional symplectic manifold, whereas the cartesian product of the state spaces is an eight dimensional symplectic manifold.  While the fibered product is an embedded submanifold of the product, it will not be a symplectic submanifold when endowed with the symplectic structure that it requires to be the state space of the given classical system. The Lagrangian setting is similar, but uses tangent bundles rather than cotangent bundles as the state spaces.  The fibered product together with its canonical projections encapsulates the physical meaning of identifying the right mass of the left spring-mass system with the left mass of the right spring-mass system.  Both Dazord in \cite{Daz} and Marle in \cite{Mar} had similar insights with respect to studying constrained systems, which are similar to the systems given above in the sense that the masses that connect our systems can be thought of as a geometric constraint.  In fact, Dazord explicitly uses fibered products to construct the configuration and state spaces for certain constrained systems.

Suppose that $X$, $Y$, and $Z$ are sets and $f$ and $g$ are functions that respectively map $X$ and $Y$ to the set $Z$.  Henceforth denote by $\rho_X$ and $\rho_Y$ the respective canonical projections \[\rho_X\colon X\times Y \to X\quad {\rm and}\quad \rho_Y\colon X\times Y \to Y,\] and denote by $\pi_X$ and $\pi_Y$ the respective restrictions of $\rho_X$ and $\rho_Y$ to the fibered product $X\times_Z Y$, which is the subset of $X \times Y$ consisting of all elements on which $f$ is equal to $g$.  The fibered product in the category $\Set$, whose objects are sets and whose morphisms are functions, has certain universal properties recalled in Section~\ref{sec:physsys}.  The connection between these universal properties and the construction of span categories for modeling classical mechanical systems is a central theme of the current investigation.

A span in the category $\Set$ is a pair of functions with the same source.  The fibered product together with the span $(\pi_X, \pi_Y)$ gives a prescription for composing spans in $\Set$.   B\'enabou proved in \cite{Ben} that if $\C$ is a category with pullbacks then there is a bicategory, ${\rm Span}\!\(\C\)$, whose objects, morphisms, and $2$-morphisms are the respective objects, spans, and maps of spans in $\C$.  To avoid unnecessary complications we view this bicategory as a category, a \emph{span category}, by ignoring the bicategory structure and taking isomorphism classes of spans in $\C$, to be defined in Section~\ref{sec:physsys}, as the morphisms.  Fibered products define a composition of isomorphism classes of certain spans in $\Set$ that seems strikingly similar to the way in which classical mechanical systems compose.  

We propose that open classical systems are morphisms in an appropriate span category, where composition of morphisms using pullbacks describes the composition of classical systems.  This formalization of classical mechanics should deepen our understanding of the foundations of classical mechanics and may also offer a way to automate the modeling of classical mechanical systems.   Modeling open classical mechanical systems necessitates working with spans in categories other than $\Set$, where the fibered product lacks the universal properties that it has in $\Set$.  

 It is natural to view a classical system as an isomorphism class of spans in the category of Riemannian manifolds with surjective Riemannian submersions in the Lagrangian setting, or as an isomorphism class of spans in the category of symplectic manifolds with surjective Poisson maps in the Hamiltonian setting.   However, Section~\ref{sec:physsys} demonstrates that neither of these categories has pullbacks.  Thus, the work of B\'enabou does not apply.  For this same reason, it does not appear that the work of Fong \cite{Fong,FongThesis} as corrected by Courser \cite{BC,CourserThesis} can be straightforwardly modified from its cospan setting to a span setting that is useful to the present discussion.  Derived geometry \cite{Calaque,Spiv2} would let us use homotopy pullbacks instead of pullbacks, but in some sense this is overkill: the fibered products required for the current paper will exist and be smooth manifolds; only the universality condition of a pullback fails.

Section~\ref{sec:physsys} recalls previous work required for handling this problem. Suppose that $\C$ and $\C^\prime$ are categories and $\F$ is a functor from $\C$ to $\C^\prime$.  Weisbart and Yassine defined in \cite{WY} the notion of an ``$\F$-pullback'' of a cospan in $\C$ and the ``span tightness'' of the functor $\mathcal F$.  They proved that if the functor $\mathcal F$ is span tight, then ${\rm Span}(\C, \F)$ is a category, a ``generalized span category'', whose objects are the objects in $\C$ and whose morphisms are isomorphism classes of spans in $\C$. Composition in this generalized span category is defined using $\F$-pullbacks. Generalized span categories determine the kinematical properties of open classical systems in the Hamiltonian setting and of ``free'' open systems in the Lagrangian setting---that is, systems where all the energy is kinetic.

In Section~\ref{secLHS} we introduce the notion of an ``augmented'' span, which allows us to introduce nonzero Hamiltonians and add potentials to Lagrangians.   In Section~\ref{sec:PhyMorph} we construct the augmented generalized span categories $\HamSy$ and $\LagSy$.   In the Hamiltonian setting, the augmentation determines the dynamical evolution of the system.  In the Lagrangian setting, the augmentation determines the potential for the classical system, hence its dynamics as well. The categories $\LagSy$ and $\HamSy$ provide frameworks for studying open classical systems from the Lagrangian and Hamiltonian perspectives, respectively.  Section~\ref{sec:Functor} introduces a functor $\FLag \colon \LagSy \to \HamSy$.  This functor, a version of the Legendre transformation, translates from the Lagrangian to the Hamiltonian perspective. 

In future work we hope to compare the present approach to the theory of port-Hamiltonian systems, an approach to open systems in classical mechanics widely used in engineering \cite{VanDerSchaft}.


\section{Spans and Generalized Span Categories}\label{sec:physsys}

\subsection{Spans and Span Categories} Refer to \cite{WY} for further discussion of the material presented in the following subsection.  We review for the reader's convenience some of the definitions and basic results from \cite{WY} that the current discussion requires.

A \emph{span} in a category $\C$ is a pair of morphisms in $\C$ with the same source and a \emph{cospan} in $\C$ is a pair of morphisms in $\C$ with the same target.  For any span $S$ in $\C$, write \[S = \(\sL, \sR\),\] where $\SL$, $\SR$, and $\SA$ are objects in $\C$, \[\sL\colon \SA\to \SL, \quad {\rm and} \quad \sR\colon \SA\to \SR.\] Utilize the same notation if $S$ is a cospan, but where $s_L$ and $s_R$ respectively map $\SL$ and $\SR$ to $\SA$.   For any span or cospan $S$ in $\C$, refer respectively to the objects $\SA$, $\SL$, and $\SR$ in $\C$ as the apex, left foot, and right foot of $S$.

\begin{definition}
A span $S$ in $\C$ is \emph{paired with} a cospan $C$ in $\C$ if \[\CL = \SL, \quad \CR = \SR,\quad {\rm and} \quad \cL\circ \sL=\cR\circ \sR.\] 
\end{definition}
The pairing of a span $S$ with a cospan $C$ has a diagrammatical interpretation, namely that this diagram commutes:
\begin{center}
\begin{showTikz}
\begin{tikzpicture}[->=stealth',node distance=2.25cm, auto, scale = .8]
\node(K) at (-2,0) {$\CL = \SL$};
\node(L) at (2,0) {$\CR = \SR$};
\node(M) at (0,2) {$\SA$};
\draw[->] (M) to node[yshift = 6, inner sep = 2, left] {$\sL$} (K);
\draw[->] (M) to node[yshift = 6, inner sep = 2, right] {$\sR$} (L);
\node(MM) at (0,-2) {$\CA$};
\draw[->] (K) to node[yshift = -6, inner sep = 2, left] {$\cL$}(MM);
\draw[->] (L) to node[yshift = -6, inner sep = 2, right] {$\cR$}(MM);
\end{tikzpicture}
\end{showTikz}
\begin{showPDF}
\includegraphics{Figure5.pdf}
\end{showPDF}
\end{center}

Suppose that $S$ and $Q$ are spans in $\C$ with $\SL$ equal to $\QL$ and $\SR$ equal to $\QR$. A \emph{span morphism in $\C$ from $S$ to $Q$} is a morphism $\Phi$ in $\C$ from $\SA$ to $\QA$ such that \[\sL = \qL\circ\Phi \quad {\rm and} \quad \sR = \qR\circ\Phi,\]
meaning that this diagram commutes:
\begin{center}
\begin{showTikz}
\begin{tikzpicture}[->=stealth',node distance=2.25cm, auto, scale = .8]
\node(L) at (-2,0) {$\SL = \QL$};
\node(R) at (2,0) {$\SR = \QR$};
\node(T) at (0,2) {$\SA$};
\node(B) at (0,-2) {$\QA$};
\draw[->] (T) to node[yshift = 6, inner sep = 2, left] {$\sL$} (L);
\draw[->] (T) to node[yshift = 6, inner sep = 2, right] {$\sR$} (R);
\draw[->] (B) to node[yshift = -6, inner sep = 2, left] {$\qL$} (L);
\draw[->] (B) to node[yshift = -6, inner sep = 2, right] {$\qR$} (R);
\draw[->] (T) to node[inner sep = 3, left] {$\Phi$} (B);
\end{tikzpicture}
\end{showTikz}
\begin{showPDF}
\includegraphics{Figure6.pdf}
\end{showPDF}
\end{center}
%
%
%
A \emph{span isomorphism in $\C$ from $S$ to $Q$} is a span morphism that is additionally an isomorphism.

\begin{proposition}\label{prop:spammap}
For any span isomorphism $\Phi$, the inverse $\Phi^{-1}$ is also a span isomorphism.  Furthermore, any composite of span morphisms is again a span morphism.
\end{proposition}

\begin{definition}
A span $S$ in $\C$ is a \emph{pullback} of a cospan $C$ in $\C$ if it is paired with $C$ and if for any other span $Q$ in $\C$ that is also paired with $C$ there exists a unique span morphism $\Phi$ in $\C$ from $Q$ to $S$:
\begin{center}
\begin{showTikz}
\begin{tikzpicture}[->=stealth',node distance=2.25cm, auto]
%
\node(W) at (0,0) {$\CL$};
\node(X) at (0,2) {$\SA$};
\node(Y) at (2.5,2) {$\CR$};
\node(Z) at (2.5,0) {$\CA$};
\node(R) at (-1.65,3.32) {$\QA$};
\draw[->] (X) to node[inner sep = 5, above] {$\sR$}(Y);
\draw[->] (X) to node[inner sep = 5, left] {$\sL$}(W);
\draw[->] (Y)to node[inner sep = 5, right] {$\cR$}(Z);
\draw[->] (W)to node[inner sep = 5, below] {$\cL$}(Z);
\draw[->] (R) to (X);
\node[] at (-.48, 2.9) {$\exists ! \Phi$};
\draw[->,bend right] (R) to node[inner sep = 5, below]{\phantom{$\displaystyle\int$}$\qL$\phantom{$\displaystyle\int$}}(W);
\draw[->,bend left](R)to node[inner sep = 5, above]{$\qR$}(Y);
\end{tikzpicture}
\end{showTikz}
\begin{showPDF}
\includegraphics{Figure7.pdf}
\end{showPDF}
\end{center}
%
%
%
\end{definition}

Notice that the diagram formed by pairing a span $S$ with a cospan $C$, where $S$ is a pullback of $C$, is a pullback diagram or a pullback square as often discussed in the literature.

\begin{definition}\label{def:hasPB}
A category $\C$ \emph{has pullbacks} if for any cospan $C$ in $\C$ there is a span $S$ in $\C$ that is a pullback of $C$ and $S$ is unique up to a span isomorphism in $\C$.
\end{definition}

Denote by $\Top$ the category whose objects are topological spaces and whose morphisms are continuous functions.  The categories $\Set$ and $\Top$ are examples of categories that have pullbacks, as discussed in \cite{WY}.  If $C$ is a cospan in $\Set$, then let $\rho_L$ and $\rho_R$ be the canonical projections \[\rho_L\colon \CL\times\CR \to \CL\quad {\rm and} \quad \rho_R\colon \CL\times\CR \to\CR.\]  Denote by $\SA$ the fibered product \[C_L\times_{\CA}\CR := \{(x,y)\in \CL\times\CR\colon (\cL\circ \rho_L)(x,y) = (\cR\circ \rho_R)(x,y)\}.\]   Take $\SL$ and $\SR$ to be respectively equal to $\CL$ and $\CR$, and let $\sL$ and $\sR$ be the respective restrictions of $\rho_L$ and $\rho_R$ to the set $\SA$.  The span $(\sL, \sR)$ is a pullback of $C$.  If $C$ is a cospan in $\Top$, then $S$ is again a pullback of $C$ in $\Top$, where the topology on $\SA$ is the subspace topology induced by the product topology on $\SL\times\SR$.


\subsection{The Categories $\SympSurj$ and $\Riem$} Refer to \cite{Lib} for further background on Poisson geometry.  A \emph{Poisson bracket} on a smooth manifold $M$ is an anticommutative, bilinear function from $C^\infty(M)\times C^\infty(M)$ to $C^\infty(M)$ that satisfies Leibniz's rule and the Jacobi identity.  A \emph{Poisson manifold} is the pair consisting of a smooth manifold $M$ and a Poisson bracket on $M$.  Suppose that $\(M,\{\cdot,\cdot\}_M\)$ and $\(N,\{\cdot,\cdot\}_N\)$ are Poisson manifolds.  For each $f$ in $C^{\infty}(M)$, the \emph{Poisson vector field associated to $f$} is the derivation $v_{f}$ given by \[v_f(\cdot)=\{\cdot,f\}_M.\] A smooth map $\Phi$ from $M$ to $N$ is a \emph{Poisson map} if for any $f$ and $g$ in $C^{\infty}(N)$, 
\[
\left\{f,g\right\} _{N}\circ\Phi\ =\ \left\{ f\circ\Phi,g\circ\Phi\right\} _{M}.
\]
Symplectic manifolds are the primary objects of study in Hamiltonian mechanics.  A \emph{symplectic manifold} is a pair $(M, \tform{M})$ where $M$ is a smooth (necessarily even dimensional) manifold and $\tform{M}$ is a smooth, closed, nondegenerate 2-form on $M$, a \emph{symplectic 2-form}.  Suppose that the dimension of $M$ is $2m$.  For each $x$ in $M$, there is an open set $U$ containing $x$ such that the symplectic 2-form gives rise to \emph{Darboux coordinates} $(q_i, p_i)_{i=1}^m$ on $U$, coordinates such that \[\tform{M} = \sum_{i=1}^m {\rm d}q_i\wedge{\rm d}p_i.\] The symplectic 2-form naturally distinguishes position and momentum coordinates on $M$ and induces an isomorphism $\Omega_M$ between the tangent and cotangent bundles.  Given tangent vectors $v$ and $w$ in the same fiber of $TM$, define by $\Omega_M(v)$ the covector \[\Omega_M(v) = \tform{M}(\cdot, v)\colon w\mapsto \tform{M}(w,v).\]  Since $\tform{M}$ is nondegenerate, the map $\Omega_M$ is invertible.  For each function $f$ in $C^\infty(M)$, denote by $D_f$ the \emph{symplectic gradient of $f$}, which is defined by \[D_f = \Omega^{-1}_M({\rm d}f).\]

Every symplectic manifold has a Poisson structure that it inherits from its symplectic structure in the following way.  For any symplectic manifold  $\(M, \tform{M}\)$, define a Poisson bracket $\{\cdot, \cdot\}_M$ on pairs $(f,g)$ in $C^\infty(M)\times C^\infty(M)$ by \[\{f,g\}_M=\tform{M}\!\(D_{f},D_{g}\).\]  The symplectic gradient $D_f$ is the Poisson vector field $v_f$ associated to $f$, implying that \[\{f,g\}_M=\tform{M}\!\(v_{f},v_{g}\).\] The real valued function $\Pi_M$ defined by \[\Pi_M\!\(\d f,\d g\)=\{f,g\}_M\] is a global section of $\(T^\ast M\wedge T^\ast M\)^\ast$.  The \emph{Poisson bivector of $\(M, \{\cdot, \cdot\}_M\)$} is the image of the function $\Pi_M$ under the canonical isomorphism that takes $\(T^\ast M\wedge T^\ast M\)^\ast$ to $\Lambda^{2}TM$.  To simplify notation, denote henceforth by $\Pi_M$ the Poisson bivector of $\(M, \{\cdot, \cdot\}_M\)$. Refer to \cite[p.~30]{wein} for Proposition~\ref{2:prop:PoissonbyPF} and see \cite[p.~44]{wein} for a proof of Proposition~\ref{2:prop:PoissonIsSubmersion}.

\begin{proposition}\label{2:prop:PoissonbyPF}
A smooth map $\Phi$ from $\(M, \{\cdot, \cdot\}_M\)$ to $\(N, \{\cdot, \cdot\}_N\)$ is a Poisson map if and only if \[\d \Phi\!\(\Pi_M\) = \Pi_N.\] 
\end{proposition}

\begin{proposition}\label{2:prop:PoissonIsSubmersion}
Suppose that $\(M, \{\cdot, \cdot\}_M\)$ is a Poisson manifold and $\(N, \omega_N\)$ symplectic manifold.  Every Poisson map from $M$ to $N$ is a submersion. 
\end{proposition}

Riemannian manifolds are the primary objects of study in Lagrangian mechanics.  The metric on the tangent bundle of a Riemannian manifold gives a kinetic energy associated to a particle moving in the base manifold which is the configuration space for the system, \cite[p.83-84]{Arn}.  A \emph{Riemannian submersion} $\Phi$ from a Riemannian manifold $\(M, g_M\)$ to a Riemannian manifold $\(N, g_N\)$ is a smooth submersion with the property that if $v$ and $w$ are vector fields tangent to the horizontal space $\({\rm ker}\!\({\rm d} \Phi\)\)^\perp$, then \[g_M\!\(v,w\) = g_N\!\(\d\Phi\!\(v\), \d\Phi\!\(w\)\).\]

Table \ref{table:categories} specifies the categories to be henceforth denoted by $\Dif$, $\SurjSub$, $\Riem$, and $\SympSurj$. 

\renewcommand{\arraystretch}{1.2}
\begin{table}
\begin{tabular}{|c |c |c |}
\multicolumn{3}{}{} \\
\hline
Category Name  & Objects  & Morphisms \\
\hline
 $\Dif$   &Smooth manifolds  & Smooth maps \\\hline
$\SurjSub$   &Smooth manifolds  & Surjective submersions \\\hline
$\Riem$   & Riemannian manifolds  & Surjective Riemannian submersions \\\hline
$\SympSurj$   & Symplectic manifolds  & Surjective Poisson maps \\
\hline
\end{tabular}
\bigskip
\caption{Table of Categories}\label{table:categories}
\end{table}

An example in \cite{WY} shows that the category $\SurjSub$ does not have pullbacks. Since this example involves manifolds that have trivial Riemannian and symplectic structures and mappings in the respective categories, the categories $\Riem$ and $\SympSurj$ also do not have pullbacks.  


\subsection{$\F$-Pullbacks and Span Tight Functors}

Assume henceforth that $\C$ and $\C^\prime$ are categories and that $\F$ is a functor from $\C$ to $\C^\prime$.  For any span $S$ in $\C$, denote by $\F(S)$ the span $(\F(\sL), \F(\sR))$ in $\C^\prime$.  For any cospan $C$ in $\C$, denote by $\F(C)$ the cospan $(\F(\cL), \F(\cR))$ in $\C^\prime$.

\begin{definition}
The category $\C$ \emph{has $\F$-pullbacks in} $\C^\prime$ if for any cospan $C$ in $\C$, there is a span $S$ in $\C$ that is paired with $C$ and the span $\F(S)$ is a pullback of the cospan $\F(C)$ in $\C^\prime$.  In this case, the span $S$ is an \emph{$\F$-pullback} of $C$.
\end{definition}

Note that if $\C^\prime$ is equal to $\C$ and $\F$ is the identity functor, then an $\F$-pullback is simply a pullback.

\begin{definition}
Suppose that $\C$ has $\F$-pullbacks in $\C^\prime$.  The functor $\F$ is \emph{span tight} if for any $\F$-pullbacks $S$ and $Q$ of the same cospan, the unique span isomorphism $\Phi$ from $\F(S)$ to $\F(Q)$ is $\F(\Psi)$ for some span isomorphism $\Psi$ from $S$ to $Q$.
\end{definition}

\begin{definition}
For any two spans $S$ and $Q$ in $\C$ such that $\SR$ is equal to $\QL$ and there is a span $P$ in $\C$ that is a pullback of the cospan $(\sR, \qL)$, denote by $S\circ_P Q$ the span in $\C$ given by \[S\circ_P Q = (\sL\circ \pL, \qR\circ \pR).\] The span $S\circ_P Q$ is the \emph{composite of $S$ and $Q$ along $P$}.  If $P$ is an $\F$-pullback, then the span $S\circ_P Q$ is an \emph{$\F$-pullback composite of $S$ and $Q$ along $P$}.
\end{definition}

Identify the objects of ${\rm Span}\!\(\C,\F\)$ to be the objects in $\C$ and the isomorphism classes of spans in $\C$ to be the morphisms in ${\rm Span}\!\(\C,\F\)$.  If $[S]$ is an isomorphism class of spans in ${\rm Span}\!\(\C, \F\)$, then identify $\SR$ and $\SL$ respectively to be the source and target of $[S]$.  Define composition of isomorphism classes of spans by \[\left[S^{\1}\right] \circ \left[S^{\2}\right] = \left[S^{\1} \circ_P S^{\2}\right],\] where $S^{\1} \circ_P S^{\2}$ is an $\F$-pullback composite of $S^{\1}$ and $S^{\2}$. For any object $X$ in $\C$, denote by ${\rm Id}_X$ the identity morphism from $X$ to $X$ and by $I_X$ the span $({\rm Id}_X, {\rm Id}_X)$.  Define by $[I_X]$ the identity morphism in ${\rm Span}\!\(\C, \F\)$ from $X$ to $X$.  The following theorem is the main result of \cite{WY}.

\begin{theorem}\label{thm:SpanCatIfTight}
If $\F$ is a span tight functor from $\C$ to $\C^\prime$, then ${\rm Span}\!\(\C, \F\)$ is a category.
\end{theorem}

Suppose that $X$, $Y$, and $Z$ are smooth manifolds.  Suppose further that $(f,g)$ is a cospan in $\SurjSub$ where $f$ and $g$ have respective sources $X$ and $Y$ and mutual target $Z$. Again denote by $\rho_X$ and $\rho_Y$  the respective projections from $X\times Y$ to $X$ and $Y$ and let $\pi_X$ and $\pi_Y$ be their respective restrictions to the embedded submanifold $X\times_ZY$. The article \cite{WY} proves Proposition~\ref{prop:TransverseansFiberedProductPB}.   Proposition~\ref{prop:TransverseansFiberedProductPB} and  Theorem~\ref{thm:SpanCatIfTight} together imply that ${\rm Span}(\SurjSub, \F)$ is a category, where $\F$ is the forgetful functor from $\SurjSub$ to $\Dif$.

\begin{proposition}\label{prop:TransverseansFiberedProductPB}
 The span $\(\pi_X, \pi_Y\)$ is an $\F$-pullback of $\(f,g\)$ and so $\SurjSub$ has $\F$-pullbacks.  Moreover, the functor $\F$ is span tight.
 \end{proposition}

Since we will need to work in the categories $\SympSurj$ and $\Riem$, we will need to prove a similar result for these categories. The following section will provide such a result.


\section{Lagrangian and Hamiltonian Systems}\label{secLHS}
 
The description of a Lagrangian or Hamiltonian system respectively requires not only the identification of a span in $\Riem$ or $\SympSurj$, but the additional information of a potential or a Hamiltonian, both of which are \emph{augmentations}.
 
\subsection{Systems as Isomorphism Classes of Augmented Spans} We now introduce the notion of an augmentation of a span and cospan, but in the restricted settings that are significant to the current discussion.  We will discuss augmentations in greater generality in an upcoming paper.

\begin{definition}
An \emph{augmented manifold} is a pair $\(M, F_M\)$, where $M$ is a smooth manifold and $F_M$ is a smooth real valued function defined on $M$.  The pair $\(M, F_M\)$ is a \emph{augmented Riemannian \lpar symplectic\rpar manifold} if $M$ is a Riemannian (symplectic) manifold.  Refer to $F_M$ as a \emph{potential \lpar or Hamiltonian\rpar}, denoting it by $V_M$ (or $H_M$) if $M$ is respectively a Riemannian (or symplectic) manifold.
\end{definition}

For sake of concision, denote by $\Cat$ any of the categories listed in Table \ref{table:categories}. 

\begin{definition}
An \emph{augmented \lpar co\rpar\!span} in $\Cat$ is a pair $\(S, F_S\)$, where $S$ is a (co)span in $\Cat$ and $F_S$ is a triple $\(F_{\SA}, F_{\SL}, F_{\SR}\)$ of smooth real valued functions defined respectively on $\SA$, $\SL$, and $\SR$.  If $\Cat$ is $\Riem$ (or $\SympSurj$), then the given augmented span is a \emph{Riemannian \lpar co\rpar\!span} (or \emph{Poisson \lpar co\rpar\!span}). A \emph{classical \lpar co\rpar\!span} is a (co)span that is either Riemannian or Poisson. If $\(S, F_S\)$ is a Riemannian (Poisson) span, then refer to $F_S$ as a \emph{potential \lpar or Hamiltonian\rpar} and denote it by $V_S$ (or $H_S$).
\end{definition}

The apex of a Poisson span determines the kinematical properties of the system and the mapping of the apex to its feet determines the way in which the span composes with other spans and, therefore, how components of systems compose to form more complicated systems.   The apex of a Riemannian span determines a free system and the augmentation will be a potential that determines the interactions in the system.  The fundamental object of our study should be an isomorphism class of augmented spans rather than an augmented span because composition using $\F$-pullbacks is only determined up to isomorphism.
 
\begin{definition}\label{Def:IsomorphicPhysicalSpan}

Suppose that the classical spans $\(S, F_S\)$ and $\(Q, F_Q\)$ are either both Riemannian or both Poisson and that \[\(\SL, F_{\SL}\) = \(\QL, F_{\QL}\) \quad {\rm and}\quad \(\SR, F_{\SR}\) = \(\QR, F_{\QR}\).\]  A span morphism $\Phi$ from $\SA$ to $\QA$  is \emph{compatible with} $F_S$ and $F_Q$ if $F_{\SA}$ is equal to $F_{\QA}\circ \Phi$ and is, in this case, a \emph{morphism of classical spans}. If $\Phi$ is additionally an isomorphism, then $\Phi$ is an \emph{isomorphism of classical spans} and $\(S, F_S\)$ and $\(Q, F_Q\)$ are \emph{isomorphic classical spans}. 
\end{definition}

The inverse of an isometry is again an isometry.  The inverse of a Poisson diffeomorphism is again a Poisson diffeomorphism, \cite[p. 10]{DZ}. Proposition~\ref{2:prop:InverseOfSpanIso} follows from these facts.

\begin{proposition}\label{2:prop:InverseOfSpanIso}
 Suppose that $S$ and $Q$ are spans that are either both in $\Riem$  or both in $\SympSurj$.  The inverse of any span isomorphism from $S$ to $Q$ is a span isomorphism from $Q$ to $S$.
\end{proposition}

Denote by $[S, F_S]$ the set of all classical spans that are isomorphic to the classical span $(S, F_S)$.  Together with the fact that the composite of classical span isomorphisms is again a classical span isomorphism, Proposition~\ref{2:prop:InverseOfSpanIso} implies that isomorphism of classical spans is an equivalence relation, hence the set $[S, F_S]$ is an equivalence class.  

\begin{definition}
A \emph{Lagrangian \lpar{\rm or} Hamiltonian\rpar system} is an isomorphism class of Riemannian (or Poisson) spans.  If $[S, F_S]$ is either a Hamiltonian system or a Lagrangian system, then $[S, F_S]$ is a \emph{classical system}.  Classical systems $[S, F_S]$ and $[Q, F_Q]$ are \emph{of the same type} if they are both Hamiltonian systems or both Lagrangian systems.
\end{definition}

\subsection{Paths of Motion}
Suppose that $(S, V_S)$ is a Riemannian span and $\met{\SA}$ is the Riemannian metric on $\SA$.  Denote by $\proj{\SA}$ the canonical projection from $T\SA$ to $\SA$.  Define the \emph{Lagrangian of $(S, V_S)$} on $T\SA$ to be the function $\Lag_S$, where \[\Lag_S(\nu) = \frac{1}{2}\met{\SA}(\nu,\nu) - V_{\SA}(\proj{\SA}(\nu)) \quad {\rm with}\quad\nu\in T\SA.\]

\begin{definition}
A path in the Riemannian manifold $\(\SA, \met{\SA}\)$ is a \emph{path of motion of $(S, V_S)$} if it minimizes the action integral of $\Lag_S$ under smooth variations with fixed endpoints.   
\end{definition}

Denote by $\g_{\SA}$ the function from $T\SA$ to $T^\ast\SA$ that acts on each $\nu$ in $T\SA$ by \[\g_{\SA}(\nu) = \met{\SA}(\nu, \cdot).\]   The nondegeneracy of the metric $\met{\SA}$ implies that the map $\g_{\SA}$ is invertible.  Denote by $\ginv_{\SA}$ the inverse of $\g_{\SA}$ with \[\ginv_{\SA}\colon T^\ast\SA \to T \SA \quad {\rm by}\quad \theta \mapsto \nu, \quad{\rm where}\quad \theta = \met{\SA}(\nu, \cdot)\quad \text{and}\quad (\theta, \nu) \in T^\ast\SA\times T\SA.\] Denote by ${\rm grad}_{\SA}\!\(V_{\SA}\)$ the vector field  \[{\rm grad}_{\SA}(V_{\SA}) = \ginv_{\SA}(\d V_{\SA}),\]  and by $\nabla^{\SA}$ the Levi-Civita connection on the Riemannian manifold $(\SA, \met{\SA})$.  A standard calculation shows that $\gamma$ is a path of motion  of the Riemannian span $(S, V_S)$ if and only if it satisfies \begin{align}\label{Equation:EL}\nabla^{\SA}_{\gamma^\prime}\gamma^\prime + {\rm grad}_{\SA}(V_{\SA}){\big |}_\gamma = 0,\tag{EL}\end{align} the Euler--Lagrange equations.  See \cite{Cort} for further explanation of the details in this section.

\begin{definition}
 Suppose that $(S, H_S)$ is a Poisson span.  Denote by $\{\cdot, \cdot\}_{\SA}$ the Poisson bracket associated to the symplectic form $\omega_{\SA}$ on the symplectic manifold $\SA$.  A path $\gamma$ in $\SA$ is a \emph{path of motion of $(S,H_S)$} if it is an integral curve of the the vector field $v$ where \[v = \left\{\cdot, H_{\SA}\right\}_{\SA}.\]
\end{definition}

\begin{proposition}\label{prop:pathsforclassicalspans}
  Suppose that $(S, F_S)$ and $(Q, F_Q)$ are classical spans of the same type and $\Phi$ is an isomorphism of classical spans taking $(S, F_S)$ to $(Q, F_Q)$. If $\gamma$ is a path of motion of $(S, F_S)$, then $\Phi\circ\gamma$ is a path of motion of $(Q, F_Q)$.  Furthermore, every path of motion of $(Q, F_Q)$ is the image of a path of motion of $(S, F_S)$.
\end{proposition}

\begin{proof}
{}{}{}{}{}{}
If $(S, V_S)$ and $(Q, V_Q)$ are Riemannian spans and $\Phi$ is an isomorphism from $(S, V_S)$ to $(Q, V_Q)$, then $\Phi$ is an isometry from $\SA$ to $\QA$ and $V_{\SA}$ is equal to $V_{\QA}\circ \Phi$.  Denote by $\nabla^{\SA}$ and $\nabla^{\QA}$ the respective Levi-Civita connections on $\SA$ and $\QA$.  Suppose that $p$ is an element of $\SA$ and that $X$ and $Y$ are tangent vector fields on $\SA$.  The map $\Phi$ is an isometry and so
\[{\rm d}\Phi_p\!\(\(\nabla^{\SA}_XY\)\!\(p\)\) = \nabla^{\QA}_{{\rm d}\Phi(X)}{{\rm d} \Phi(Y)}\!\(\Phi\(p\)\) \quad {\rm and}\quad {\rm d}\Phi\!\({\rm grad}_{\SA}\!\(V_{\QA}\circ \Phi\)\) = {\rm grad}_{\QA}\!\(V_{\QA}\).\] If $\gamma$ is a path of motion of $(S, V_S)$, then $\Phi\circ\gamma$ is a curve in $\QA$ and
\begin{align*}
\nabla^{\QA}_{\(\Phi\circ \gamma\)^\prime}\!\(\Phi\circ\gamma\)^\prime + {\rm grad}_{\QA}\!\(V_{\QA}\){\big |}_{\Phi\circ\gamma} &= \nabla^{\QA}_{{\rm d}\Phi(\gamma^\prime)}\!\({\rm d}\Phi\!\(\gamma^\prime\)\) + {\rm grad}_{\QA}\!\(V_{\QA}\){\big |}_{\Phi\circ\gamma}\\&= {\rm d}\Phi_p\!\(\nabla^{\SA}_{\gamma^\prime}\!\(\gamma^\prime\) + {\rm grad}_{\SA}\!\(V_{\SA}\){\big |}_{\gamma}\)\\ &= {\rm d}\Phi_p\!\(0\) = 0,
\end{align*}
where the fact that $\gamma$ satisfies \eqref{Equation:EL} in $\SA$ implies the penultimate equality.  The path $\Phi\circ \gamma$ is therefore a path of motion of $(Q, V_Q)$.

If $(S, H_S)$ and $(Q, H_Q)$ are Poisson spans and $\Phi$ is an isomorphism from $(S, H_S)$ to $(Q, H_Q)$, then $\Phi$ is a Poisson diffeomorphism from $\SA$ to $\QA$ and $H_{\SA}$ is equal to $H_{\QA}\circ \Phi$.  The curve $\gamma$ is path of motion of $(S, H_S)$ if and only if it is an integral curve of the vector field $\left\{\cdot, H_{\SA}\right\}$.  Suppose that $\alpha$ and $\beta$ are smooth functions on $\QA$.  Since $\Phi$ is Poisson, \[{\rm d}\Phi\!\(\left\{\cdot, \alpha\circ \Phi\right\}_{\SA}\)(\beta) = \left\{\cdot, \alpha\circ \Phi\right\}_{\SA}(\beta\circ \Phi) = \left\{\beta\circ \Phi, \alpha\circ \Phi\right\}_{\SA} = \left\{\beta,\alpha\right\}_{\QA}\] and so  
\begin{align*}
\(\Phi \circ \gamma\)^\prime &= {\rm d}\Phi{\big |}_{\gamma}\!\(\left\{\cdot, H_{\SA}\right\}_{\SA}\) \\&= {\rm d}\Phi{\big |}_{\gamma}\!\(\left\{\cdot, H_{\QA}\circ \Phi\right\}_{\SA}\) =  \left\{\cdot, H_{\QA}\right\}_{\QA}{\big |}_{\Phi\circ\gamma}.
\end{align*}%
The curve $\Phi \circ \gamma$ is, therefore, a path of motion of $(Q, H_Q)$.

In both the Riemannian and Poisson settings, the map $\Phi^{-1}$ is also an isomorphism of classical spans and so every path of motion of $(Q, F_Q)$ is the image of a path of motion of $(S, F_S)$.
{}{}{}{}{}{}{}
\end{proof}

\subsection{$\F$-Pullbacks of $\SympSurj$ and $\Riem$ in $\Dif$} Proposition~\ref{prop:pathsforclassicalspans} implies that an isomorphism class of classical spans determines the dynamics of a classical system.  Composing such isomorphism classes requires both the existence of $\F$-pullbacks in these categories, where $\F$ is an appropriate forgetful functor into $\Dif$, as well as the span tightness of the functor $\F$. 

Suppose $X$ is a symplectic manifold.  The Poisson bivector $\Pi_X$ of $X$ induces a map $\widetilde{\Pi}_X$ from $T^\ast X$ to $TX$ that takes any $\eta$ in $T^\ast X$ to the vector field $\widetilde{\Pi}_X(\eta)$ with the property that for any $\nu$ in $T^\ast X$, \[\nu\big(\widetilde{\Pi}_X(\eta)\big) =  \Pi_X(\eta, \nu).\] Since $X$ is symplectic, the map $\widetilde{\Pi}_X$ is an isomorphism \cite[p.\;17]{wein}.  This isomorphism gives a way to pull back vector fields by surjective Poisson maps, a fact that, along with Proposition~\ref{2:prop:PoissonIsSubmersion}, is critical to the proof of Theorem~\ref{thm:WeinSplit}. Theorem~\ref{thm:WeinSplit} establishes the existence of a local splitting of the tangent space of a symplectic manifold by a local foliation given by the inverse image of a surjective Poisson map.

\newcommand{\zzz}{Z}
\newcommand{\xxx}{X}
\newcommand{\dimxxx}{\ell}
\newcommand{\dimzzz}{n}
\newcommand{\tempV}{V^\prime}
\newcommand{\finalV}{V}

\begin{theorem}\label{thm:WeinSplit}
 Suppose that $\xxx$ and $\zzz$ are symplectic manifolds with respective dimensions $2\dimxxx$ and $2\dimzzz$ and that $f$ is a surjective Poisson map from $\xxx$ to $\zzz$.  Given any $z$ in $\zzz$ and a choice of Darboux coordinates $(q_{i}^\zzz, p_{i}^\zzz)_{i=1}^{\dimzzz}$ on an open set $U$ containing $z$, and given any $x$ in $\xxx$ with $f(x)$ equal to $z$, there exist Darboux coordinates $(q_{i}^\xxx,p_{i}^\xxx)_{i=1}^{\dimxxx}$ on an open set $\finalV$ containing $x$ such that for any $i$ in $\{1, \dots, \dimzzz\}$, \[q_i^\xxx=q_i^\zzz\circ f \quad {\rm and} \quad p_i^\xxx=p_i^\zzz\circ f.\]
\end{theorem}

\newcommand{\alphaa}{\eta}
\newcommand{\betaa}{\zeta}
\begin{proof}

Suppose that $x_0$ is in $\xxx$, that $U$ is an open set containing $f(x_0)$, and that $(q_{i}^\zzz, p_{i}^\zzz)_{i=1}^{\dimzzz}$ is a Darboux coordinate system on $U$.  Proposition~\ref{2:prop:PoissonIsSubmersion} guarantees that $f$ is a surjective submersion, hence it is an open map and so there is an open set $\tempV$ containing $x_0$ with a Darboux coordinate system $(q_{i}^\xxx,p_{i}^\xxx)_{i=1}^{\dimxxx}$ such that $f(\tempV)$ is an open subset of $U$. Denote by $\mathcal{H}$ the set of all vector fields $v$ on $f(\tempV)$ for which there is some $\alpha$ in $C^{\infty}(f(\tempV))$ such that for any $\beta$ in $C^{\infty}(f(\tempV))$, \[v(\beta)=\{\beta,\alpha\}_{\zzz}.\]  Denote such a vector field by $v_\alpha$. Denote by $f^{\ast}(\mathcal{H})$ the set of all vector fields $w$ on $\tempV$ for which there is an $\alpha$ in $C^\infty(f(\tempV))$ such that for any $h$ in $C^\infty(\tempV)$, \[w(h)=\{h,\alpha\circ f\}_{\xxx}.\] Denote such a vector field by $w_\alpha$.  For any $x$ in $\tempV$ and any $z$ in $f(\tempV)$, denote respectively by $f^{\ast}(\mathcal{H})(x)$ and $\mathcal{H}(z)$ the set of all vector fields in $f^{\ast}(\mathcal{H})$ evaluated at $x$ and the set of all vector fields in $\mathcal{H}$ evaluated at $z$. The bilinearity of the bracket implies that $\mathcal{H}(z)$ and $f^{\ast}(\mathcal{H})(x)$ are vectors spaces.  Since \[v_{-q^{\zzz}_{i}} = \frac{\partial}{\partial{p^{\zzz}_{i}}} \quad \textrm{and}\quad v_{p^{\zzz}_{i}} = \frac{\partial}{\partial{q^{\zzz}_{i}}},\] for any $z$ in $f(\tempV)$, the vector space $\mathcal{H}(z)$ spans $T_{z}(U)$.

Let $F$ be the function \[F\colon \mathcal{H}\to f^{\ast}(\mathcal{H}) \quad \textrm{by}\quad F(v_\alpha)=w_\alpha.\]  The fact that $f$ is Poisson implies that
\begin{align*}
{\rm d}f(w_\alpha)(\beta)&=w_\alpha(\beta\circ f)\\
&=\{\beta\circ f,\alpha\circ f\}_{\xxx}\\
&=\{\beta,\alpha\}_{\zzz}=v_{\alpha}(\beta), 
\end{align*} 
and so
 \[{\rm d}f(F(v_{\alpha}))=v_{\alpha}.\]  Similarly, for any $w_{\alpha}$ in $f^{\ast}(\mathcal{H})$, \[F({\rm d}f(w_{\alpha}))=F(v_{\alpha})=w_{\alpha}.\]  The maps $F$ and ${\rm d}f|_{\mathcal{H}}$ are therefore inverses of each other and so for each $x$ in $\tempV$, the vector spaces $\mathcal{H}(f(x))$ and $f^{\ast}(\mathcal{H})(x)$ are isomorphic.  Both of these vector spaces have the same dimension as $\zzz$.

For any $w_{\alpha}$ and $w_{\alpha^\prime}$ in $f^{\ast}(\mathcal{H})$, the Jacobi identity implies that
\begin{align*}
[w_{\alpha},w_{\alpha^\prime}]_{T\xxx}&=w_{\alpha}(w_{\alpha^\prime}(\beta))-w_{\alpha^\prime}(w_{\alpha}(\beta))\\
&=\{w_{\alpha^\prime}(\beta),\alpha\circ f\}_{\xxx}-\{w_{\alpha}(\beta),\alpha^\prime\circ f\}_{\xxx}\\
&=\{\{\beta\circ f,\alpha^\prime\circ f\}_{\xxx},\alpha\circ f\}_{\xxx}-\{\{\beta\circ f,\alpha\circ f\}_{\xxx},\alpha^\prime\circ f\}_{\xxx}\\
&=\{\beta,\{\alpha^\prime\circ f,\alpha\circ f\}_{\xxx}\}_{\xxx}=w_{\{\alpha,\alpha^\prime\}}(\beta),
\end{align*} 
and so the space of vector fields $f^{\ast}(\mathcal{H})$ is closed under the bracket $[\cdot,\cdot]_{T\xxx}$ on $T\xxx$.  Frobenius' Theorem for involutive distributions implies that for any $x$ in $\tempV$ there is a submanifold $W$ of $\tempV$ such that $f^{\ast}(\mathcal{H})(x)$ is the tangent space $T_{x}W$. Since \[f^{\ast}(\mathcal{H})(x)\cap \textrm{ker}({\rm d}f\big|_{x})=\{0\},\] the rank-nullity theorem implies that \[T_{x}\tempV=f^{\ast}(\mathcal{H})(x)\oplus \textrm{ker}({\rm d}f\big|_{x}).\]

Define the function $g$ from $W$ to $\zzz$ to be the restriction of $f$ to the submanifold $W$.  The form $g^{\ast}(\omega_{\zzz})$ is a closed 2-form on $W$ as it is the pullback of the closed 2-form $\omega_{\zzz}$ restricted to $f(\tempV)$.  Suppose that there is a $v$ in $TW$ such that for all $w$ in $TW$, $g^{\ast}(\omega_{\zzz})(v,w)$ is equal to zero.  In this case, \[0 = g^{\ast}(\omega_{\zzz})(v,w)=\omega_{\zzz}({\rm d}g(v),{\rm d}g(w)),\] and so \[\omega_{\zzz}({\rm d}g(v),\cdot)=0\] since ${\rm d}g\big|_{x}$ is surjective at each point $x$ of $W$.  Nondegeneracy of $\omega_{\zzz}$ implies that ${\rm d}g(v)$ is equal to zero and the injectivity of ${\rm d}g$ further implies that $v$ is equal to zero.  The form  $g^{\ast}(\omega_{\zzz})$ is, therefore, a symplectic form on $W$.

For any $(\alphaa, \betaa)$ in $C^\infty(\tempV)\times C^\infty(\tempV)$, %
\begin{align}\label{eq:restrictoWxW}
f^{\ast}(\omega_{\zzz})\big|_{x}(w_\alphaa,w_\betaa)&=\omega_{\zzz}({\rm d}f(w_\alphaa),{\rm d}f(w_\betaa))|_{f(x)}\notag\\
&=\omega_{\zzz}(v_\alphaa,v_\betaa)\big|_{f(x)}\notag\\
&=\{\alphaa,\betaa\}_{\zzz}\big|_{f(x)}\notag\\
&=\{\alphaa\circ f,\betaa\circ f\}_{\xxx}\big|_{x} =\omega_{\xxx}(w_\alphaa,w_\betaa)\big|_{x},
\end{align} where the assumption that $f$ is Poisson implies the penultimate equality.  The pullback $f^{\ast}(\omega_{\zzz})$ is therefore the restriction of $\omega_{\xxx}$ to $TW\times TW$. The manifold $W$ is an embedded symplectic submanifold of $\tempV$ and so \cite[p.124, Exercise 3.38]{Mcduff} implies that there is an open set $\finalV$ of $\tempV$ that contains $x_0$ and a Darboux coordinate system $(q^{\xxx}_{i},p^{\xxx}_{i})_{i=1}^{\dimxxx}$ on $\finalV$ such that for any $x$ in $\finalV$ and $i$ strictly larger than $\dimzzz$, \[q^{\xxx}_{i}(x)=p^{\xxx}_{i}(x)=0.\] %
Define \[\omega_A = \sum_{i=1}^{\dimzzz}{{\rm d}q^{\xxx}_{i}\wedge {\rm d}p^{\xxx}_{i}} \quad {\rm and}\quad \omega_B = \sum_{i=\dimzzz+1}^{\dimxxx}{{\rm d}q^{\xxx}_{i}\wedge {\rm d}p^{\xxx}_{i}},\] so that in the open set $\finalV$, $\omega_\xxx$ is equal to the sum of $\omega_A$ and $\omega_B$.  The form $\omega_B$ is the restriction of $\omega_\xxx$ to $(TW\times TW)\cap (T\finalV\times T\finalV)$ and so \eqref{eq:restrictoWxW} implies that $\omega_B$ is equal to $f^\ast(\omega_\xxx)$.  Furthermore, for any $\theta$ in $C^\infty(U)$,
\begin{align*}
 (f^{\ast}({\rm d}q^{\zzz}_{i}))(w_{\theta})|_{x}&={\rm d}q^{\zzz}_{i}({\rm d}f(w_{\theta}))\big|_{x}\\
 &={\rm d}q^{\zzz}_{i}(v_{\theta})\big|_{f(x)}\\
 &=v_{\theta}(q^\zzz_i)\big|_{f(x)}\\
 &=\{q^\zzz_i, \theta\}_\zzz\big|_{f(x)}\\
 &=\{q^\zzz_i\circ f, \theta\circ f\}_\xxx\big|_{x}
={\rm d}(q^\zzz_i\circ f)w_\theta\big|_{x}.
 \end{align*}
Every element of $TW$ is of the form $w_\theta$ for some $\theta$ in $C^\infty(U)$, implying that \begin{equation}\label{eq:coordexpressionforwz}f^{\ast}({\rm d}q^{\zzz}_{i})={\rm d}(q^{\zzz}_{i}\circ f)\quad \text{and} \quad f^{\ast}({\rm d}p^{\zzz}_{i})={\rm d}(p^{\zzz}_{i}\circ f).\end{equation}  Use \eqref{eq:coordexpressionforwz} together with the coordinate representation of $\omega_Z$ to obtain the equality \[f^{\ast}(\omega_{\zzz})=\sum_{i=1}^{\dimzzz}{{\rm d}(q^{\zzz}_{i}\circ f)\wedge {\rm d}(p^{\zzz}_{i}\circ f)},\] that implies that in the open set $\finalV$, \[\omega_{\xxx}=\sum_{i=1}^{\dimzzz}{{\rm d}(q^{\zzz}_{i}\circ f)\wedge {\rm d}(p^{\zzz}_{i}\circ f)}+\sum_{i=\dimzzz+1}^{\dimxxx}{{\rm d}q^{\xxx}_{i}\wedge {\rm d}p^{\xxx}_{i}}.\]  The coordinate system $\phi$ on $\finalV$ given by \[\phi = (q^{\zzz}_{1}\circ f,p^{\zzz}_{1}\circ f,\dots,q^{\zzz}_{\dimzzz}\circ f,p^{\zzz}_{\dimzzz}\circ f,q^{\xxx}_{\dimzzz+1},p^{\xxx}_{\dimzzz+1},\dots,q^{\xxx}_{\dimxxx},p^{\xxx}_{\dimxxx})\] is, therefore, a Darboux coordinate system on $\finalV$.%
 \end{proof}

Denote by $\pi_Z$ the map \[\pi_Z = f\circ \pi_X = g\circ \pi_Y,\] where $\pi_X$ and $\pi_Y$ are the projections from $X\times_ZY$ to $Z$.  More generally, for any span $Q$ that is paired with a cospan $(f,g)$, define by $\qM$ the map \[\qM = f\circ \qL = g\circ\qR.\]

 \begin{theorem}\label{Thm:XYPoissonExistence}
 Suppose that $\(f, g\)$ is a cospan in $\SympSurj$ with \[f\colon X\to Z\quad {\rm and}\quad g\colon Y\to Z,\] with $2\ell$, $2m$, $2n$ the respective dimensions of $X$, $Y$, and $Z$, and suppose that $\omega_X$, $\omega_Y$, and $\omega_Z$ are the respective symplectic forms on $X$, $Y$, and $Z$.  Suppose that $Q$ is a span in $\SympSurj$ that is paired with $(f,g)$ and suppose that $\QA$ has dimension $2(\ell + m - n)$.  The 2-form $\omega_{\QA}$, given by \[\omega_{\QA} = \qL^\ast\!\(\omega_X\) + \qR^\ast\!\(\omega_Y\) - \qM^\ast\!\(\omega_Z\),\] is the symplectic form on $\QA$.  Moreover, the 2-form $\omega$, given by \[\omega = \pi_X^\ast\!\(\omega_X\) + \pi_Y^\ast\!\(\omega_Y\) - \pi_Z^\ast\!\(\omega_Z\)\] is the unique symplectic form on $X\times_{Z}Y$ with the property that $\(\pi_X, \pi_Y\)$ is paired with $\(f, g\)$.
 \end{theorem}
 
 \newcommand{\xlbl}{X}
 \newcommand{\ylbl}{Y}
  \newcommand{\zlbl}{Z}
   
\begin{proof}
{}{}{}{}{}{}

Suppose that $a$ is in $\QA$.  Since $Z$ is a symplectic manifold, there is on some open set $U_Z$ containing $\qM\!\(a\)$ a Darboux coordinate system $\Psi^{\zlbl}$ with \[\Psi^{\zlbl} = \(q^{\zlbl}_k, p^{\zlbl}_k\)_{k\in\{1,\dots, n\}} \colon U_Z \to \mathbb R^{2n}.\]  Since $\qM\!\(a\)$ is equal to $f\!\(\qL\!\(a\)\)$, Theorem~\ref{thm:WeinSplit} implies that there is an open set $U_X$ containing $\qL\!\(a\)$ and a Darboux coordinate system $\Psi^{\xlbl}$ on $U_{\xlbl}$ with \[\Psi^{\xlbl} = {\(q^{\xlbl}_i, p^{\xlbl}_i, q^{\zlbl}_k\circ f, p^{\zlbl}_k\circ f\)_{\;\substack{\mathllap{i} \in \mathrlap{\{1,\dots, \ell-n\}} \\ {\mathllap{k} \in \mathrlap{\{1,\dots, n\}} }}}}\qquad\quad\;\;\colon U_{\xlbl} \to \mathbb R^{2\ell}.\]  Similarly, there is an open set $U_{\ylbl}$ containing $\qR\!\(a\)$ and a Darboux coordinate system $\Psi^{\ylbl}$ on $U_{\ylbl}$ with \[\Psi^{\ylbl} = {\(q^{\ylbl}_j, p^{\ylbl}_j, q^{\zlbl}_k\circ g, p^{\zlbl}_k\circ g\)_{\;\substack{\mathllap{j} \in \mathrlap{\{1,\dots, m-n\}} \\ {\mathllap{k} \in \mathrlap{\{1,\dots, n\}} }}}}\qquad\quad\;\;\;\colon U_{\ylbl} \to \mathbb R^{2m}.\]

For each $k$ in $\{1, \dots, n\}$, the equality of $f\circ \qL$ and $ g\circ \qR$ implies that \[q^{\zlbl}_k \circ f\circ \qL = q^{\zlbl}_k \circ g\circ \qR = q^{\zlbl}_k \circ \qM\quad {\rm and}\quad p^{\zlbl}_k \circ f\circ \qL = p^{\zlbl}_k \circ g\circ \qR = p^{\zlbl}_k \circ \qM.\] Furthermore, there is an open set  $U$ containing $a$ with the property that $\qL\!\(U\)$ and $\qR\!\(U\)$ are, respectively, subsets of $U_{\xlbl}$ and $U_{\ylbl}$.  Denote respectively by $\tilde{q}^{\xlbl}_i$, $\tilde{p}^{\xlbl}_i$, $\tilde{q}^{\ylbl}_j$, $\tilde{p}^{\ylbl}_j$, $\tilde{q}^{\zlbl}_k$, $\tilde{p}^{\zlbl}_k$ the functions $q^{\xlbl}_i\circ\qL$, $p^{\xlbl}_i\circ\qL$, $q^{\ylbl}_j\circ\qR$, $p^{\ylbl}_j\circ\qR$, $q^{\zlbl}_k\circ\qM$, and $p^{\zlbl}_k\circ\qM$ acting on $\QA$.  The map $\Psi$ given by \[\Psi = \(\tilde{q}^{\xlbl}_i, \tilde{p}^{\xlbl}_i, \tilde{q}^{\ylbl}_j, \tilde{p}^{\ylbl}_j, \tilde{q}^{\zlbl}_k, \tilde{p}^{\zlbl}_k\)_{\;\substack{\mathllap{i} \in \mathrlap{\{1,\dots, \ell-n\}} \\\mathllap{j} \in \mathrlap{\{1,\dots, m-n\}} \\ {\mathllap{k} \in \mathrlap{\{1,\dots, n\}} }}} \qquad\quad\;\;\colon U \to \mathbb R^{2\(\ell+m-n\)}\] is a homeomorphism from $U$ to an open subset of $\mathbb R^{2(\ell+m-n)}$ and hence a coordinate system on $U$ that is a Darboux coordinate system.  The 2-form $\omega_{\QA}$ is therefore the form \[\omega_{\QA} = \sum_{i=1}^{\ell-n}{\rm d}\tilde{q}^{\xlbl}_i\wedge{\rm d}\tilde{p}^{\xlbl}_i + \sum_{j=1}^{m-n}{\rm d}\tilde{q}^{\ylbl}_j\wedge{\rm d}\tilde{p}^{\ylbl}_j + \sum_{k=1}^{n}{\rm d}\tilde{q}^{\zlbl}_k\wedge{\rm d}\tilde{p}^{\zlbl}_k,\] proving that if there is a span $Q$ with the given properties, then the symplectic form on $\QA$ is determined by the cospan $(f,g)$.  It does not, however, prove that there is such a span.

Proposition~3.6 of \cite{WY} implies that $X\times_ZY$ is a smooth manifold of dimension $2(\ell+m-n)$.  Suppose $v$ is in $T_{a}\!\(X\times_{Z} Y\)$ and for any $w$ in $T_{a}\!\(X\times_{Z} Y\)$, $\omega(v,w)$ is zero.  There are coefficients $a^{i},b^{i},c^{j},e^{j}, s^{k},t^{k}$ such that, using Einstein summation convention, \[v=a^{i}\partial{\tilde{q}^{\xlbl}_{i}}+b^{i}\partial{\tilde{p}^{\xlbl}_{i}}+c^{j}\partial{\tilde{q}^{\ylbl}_{j}}+e^{j}\partial\tilde{p}^{\ylbl}_j+s^{k}\partial\tilde{q}^{\zlbl}_k+t^{k}\partial\tilde{p}^{\zlbl}_k.\]  For a fixed $i$, \[-\omega(v,\partial{\tilde{q}^{\xlbl}_{i}})= b^{i}=0.\]  A similar calculation shows that all of the given coefficients are zero, implying that $v$ is equal to zero and so $\omega$ is nondegenerate.  The form $\omega$ is the sum of pullbacks of smooth closed forms, and so smooth and closed itself, hence symplectic.  The construction of $\omega$ ensures that the smooth surjections $\pi_X$ and $\pi_Y$ are Poisson maps on the symplectic manifold $(X\times_ZY, \omega)$, hence $(\pi_X, \pi_Y)$ is paired with $(f,g)$.
{}{}{}{}{}{}{}
 \end{proof}

 \begin{theorem}\label{Thm:XYRiemannianExistencez}
 Suppose that $\(f, g\)$ is a cospan in $\Riem$ with \[f\colon X\to Z\quad {\rm and}\quad g\colon Y\to Z\] and that $g_X$, $g_Y$, and $g_Z$ are the metric tensors on $X$, $Y$, and $Z$, respectively.  The tensor $g_{X\times_ZY}$, given by \[g_{X\times_ZY} = \pi_X^\ast\!\(g_X\) + \pi_Y^\ast\!\(g_Y\) - \pi_Z^\ast\!\(g_Z\),\]  is the unique metric tensor on $X\times_{Z}Y$ such that the span $\(\pi_X, \pi_Y\)$ is paired with $\(f, g\)$. 
 \end{theorem}

\begin{proof}
{}{}{}{}{}{}
Since every surjective Riemannian submersion is a surjective submersion, the fibered product $X\times_ZY$ is a smooth manifold.   If $g_{X\times_ZY}$ is positive definite, then $\(X\times_ZY, g_{X\times_ZY}\)$ is a Riemannian manifold since $g_{X\times_ZY}$ is a symmetric tensor as a sum of pullbacks of symmetric tensors.  It suffices to show that $g_{X\times_ZY}$ is nondegenerate.  

Follow the proof of Theorem~\ref{Thm:XYPoissonExistence}, using the splitting of the tangent spaces \[TX = \({\rm ker}\!\(\d f\)\)^\perp\oplus\({\rm ker}\!\(\d f\)\) \quad {\rm and}\quad TY = \({\rm ker}\!\(\d g\)\)^\perp\oplus\({\rm ker}\!\(\d g\)\)\] rather than the previous appeal to Theorem~\ref{thm:WeinSplit} to obtain an expression for $g_{X\times_ZY}$ in local coordinates.  Together with this local coordinate representation of $g_{X\times_ZY}$, the fact that the maps $\pi_X$, $\pi_Y$ and $\pi_Z$ are surjective Riemannian submersions implies that $g_{X\times_ZY}$ is nondegenerate.  The proof is similar to the proof of Theorem~\ref{Thm:XYPoissonExistence} and so the details are left to the reader to verify.
{}{}{}{}{}{}{}
\end{proof}

Note that the symplectic form on $X\times_ZY$ in Theorem~\ref{Thm:XYPoissonExistence} is not the pullback by the inclusion map of the symplectic form on $X\times Y$ to the manifold $X\times_ZY$.  While the pullback form is symplectic, the span $(\pi_X, \pi_Y)$ will no longer be a span in $\SympSurj$ when $X\times_ZY$ is endowed instead with the pullback form.  The analogous statements about the potential choices for the metric tensor are true in the Riemannian setting.

\begin{theorem}\label{thm:RDFspanTight}
The forgetful functors from $\SympSurj$ to $\Dif$ and from $\Riem$ to $\Dif$ are span tight.
\end{theorem}

\begin{proof}
{}{}{}{}{}{}
Suppose that $\F$ is the forgetful functor from $\SympSurj$ to $\Dif$.  Since every morphism in $\SympSurj$ is a surjective submersion, the functor $\F$ maps $\SympSurj$ to the subcategory $\SurjSub$ of $\Dif$.  If $\(f, g\)$ is a cospan in $\SympSurj$, and $\pi_X$ and $\pi_Y$ are, as defined above, the respective projections from $X\times_ZY$ to $X$ and $Y$, then Proposition~\ref{prop:TransverseansFiberedProductPB} implies that $\(\F\!\(\pi_X\), \F\!\(\pi_Y\)\)$ is a span in $\Dif$ that is a pullback of the cospan $\(\F\!\(f\), \F\!\(g\)\)$.  Therefore, $\SympSurj$ has $\F$-pullbacks in $\Dif$.  Suppose now that $Q$ is a span in $\SympSurj$ that is also an $\F$-pullback of $\(f, g\)$.  In this case, the span $\F(Q)$ is a span in $\Dif$ that is a pullback of $\(\F\!\(f\), \F\!\(g\)\)$ and so there is a span diffeomorphism $\Phi$ from $\F(Q)$ to $\F\!\(X\times_ZY\)$.  Since $\Phi$ is a span morphism, \begin{equation}\label{eqs:spaniso1}\F(\qL)\circ\Phi^{-1} = \F(\pi_X), \quad \F(\qR)\circ\Phi^{-1} = \F(\pi_Y), \quad {\rm and}\quad \F(f)\circ\F(\qL)\circ\Phi^{-1} = \F(\pi_Z).\end{equation}  Denote respectively by $\omega$, $\omega_X$, $\omega_Y$, and $\omega_Z$ the symplectic forms on $X\times_Z Y$, $X$, $Y$, and $Z$.  The  equalities of \eqref{eqs:spaniso1} imply that %
\begin{align*}
\omega &= \F\!\(\pi_X\)^\ast\!\(\omega_X\) + \F\!\(\pi_Y\)^\ast\!\(\omega_Y\)- \F\!\(\pi_Z\)^\ast\!\(\omega_Z\)\\
&= \(\F(\qL)\circ \Phi^{-1}\)^\ast\!\(\omega_X\) + \(\F(\qR)\circ \Phi^{-1}\)^\ast\!\(\omega_Y\)- \(\F(f)\circ\F(\qL)\circ \Phi^{-1}\)^\ast\!\(\omega_Z\)\\
&= \(\Phi^{-1}\)^\ast\Big(\F(\qL)^\ast\(\omega_X\) + \F(\qR)^\ast\!\(\omega_Y\)- \(\F(f)\circ\F(\qL)\)^\ast\!\(\omega_Z\)\Big)\\
&=\(\Phi^{-1}\)^\ast\!\(\tform{\QA}\),
\end{align*}
where $\tform{\QA}$ is the unique 2-form on $\QA$ such that $Q$ is paired with $(f,g)$.  Let $\Psi$ be the map from $(\QA, \tform{\QA})$ to $(X\times_ZY, \omega)$ that acts as $\Phi$ on the underlying manifolds.  The map $\Psi$ is, therefore, a diffeomorphism and $\Psi^{-1}$ is a symplectic map, hence $\Psi$ is a symplectomorphism.  Since every symplectomorphism is a Poisson diffeomorphism, $\Psi$ is an isomorphism in the category $\SympSurj$ with $\F(\Psi)$ equal to $\Phi$, \cite[p. 195]{AM}.  

A similar argument proves the theorem in the case of $\Riem$.
{}{}{}{}{}{}{}
\end{proof}

\begin{corollary}
If $\F$ is the forgetful functor from $\SympSurj$ to $\Dif$ \lpar resp. $\Riem$ to $\Dif$ \rpar, then ${\rm Span}(\SympSurj, \F)$ \lpar resp. ${\rm Span}(\Riem, \F)$\rpar is a category.
\end{corollary}

While Theorems~\ref{thm:SpanCatIfTight} and \ref{thm:RDFspanTight} imply that  ${\rm Span}(\SympSurj, \F)$ and ${\rm Span}(\Riem, \F)$ are categories, where $\F$ is the appropriate forgetful functor into $\Dif$, to show that classical systems are morphisms of a category requires additional verifications.  The next section provides the necessary verifications.



\section{Classical Systems as Morphisms}\label{sec:PhyMorph}
This section constructs the categories $\LagSy$ and $\HamSy$, whose objects are respectively augmented Riemannian manifolds or augmented symplectic manifolds and whose morphisms are isomorphism classes of the classical spans appropriate to the given category.  

\subsection{The Categories $\HamSy$ and $\LagSy$}

\begin{definition}
The classical system $[S,F_S]$ is \emph{composable} with the classical system $[Q, F_Q]$ if:
\begin{enumerate}
\item[(i)] both are classical systems of the same type; 
\item[(ii)] if $\(S, F_S\)$ and $\(Q, F_Q\)$ are respective representatives of the equivalence classes $[S, F_S]$ and $[Q, F_Q]$, then $\(\SR, F_{\SR}\)$ is equal to $\(\QL, F_{\QL}\)$. 
\end{enumerate}
\end{definition}

Assume below that the classical system $\left[S,F_S\right]$ is composable with $\left[Q,F_Q\right]$, and $\(S,F_S\)$ and $\(Q,F_Q\)$ are, respectively, representatives of $\left[S,F_S\right]$ and $\left[Q,F_Q\right]$.  To simplify notation, let  \[\SA = X,\; \SL = V,\; \SR = \QL = Z,\; \QA = Y,\; {\rm and}\; \QR = W.\]  Again denote by $X\times_Z Y$ the fibered product and by $\pi_X$, $\pi_Y$, and $\pi_Z$ the respective projections to $X$, $Y$, and $Z$.  Define by $[S,F_S]\circ[Q,F_Q]$ the augmented span given by \begin{align*}[S,F_S]\circ [Q, F_Q] &= \left[\(\sL\circ\pi_X, \qR\circ\pi_Y\), F_{S\circ Q}\right],\end{align*} where \[F_{S\circ Q} = \(F_X\circ\pi_X + F_Y\circ\pi_Y - F_Z\circ\pi_Z, F_V, F_W\).\]
 
\begin{theorem}
The Hamiltonian systems are the morphisms in a category, $\HamSy$, whose objects are augmented symplectic manifolds.  The Lagrangian systems are the morphisms in a category, $\LagSy$, whose objects are augmented Riemannian manifolds.
\end{theorem}

\begin{proof}
{}{}{}{}{}{}
To prove the theorem, it suffices to show that: (i) composition of morphisms in $\HamSy$ and in $\LagSy$ is well defined; (ii) both $\HamSy$ and $\LagSy$ have left and right unit laws; and (iii) composition of morphisms in $\HamSy$ and in $\LagSy$ is associative.  Since ${\rm Span}(\Riem, \F)$ and ${\rm Span}(\SympSurj, \F)$ are categories, to show that $\HamSy$ and $\LagSy$ are categories, it suffices to show that the augmentations are compatible with the various span isomorphisms that arise in defining the categories ${\rm Span}(\Riem, \F)$ and ${\rm Span}(\SympSurj, \F)$. Suppose that $\left[S,F_{S}\right]$ and $\left[Q,F_{Q}\right]$ are both classical systems and denote by $\F$ the appropriate forgetful functor from either $\SympSurj$ or $\Riem$ to $\Dif$. 

(i) Suppose that $\left[S^{\prime},F_{S^\prime}\right]$ is equal to $\left[S,F_{S}\right]$ and that $\alpha$ is an isomorphism of augmented spans with \[\alpha \colon X = \SA \to \SA^\prime.\]  Suppose that $\left[Q^{\prime},F_{Q^\prime}\right]$ is equal to $\left[Q,F_{Q}\right]$ and that $\beta$ is an isomorphism of augmented spans with \[\beta \colon Y = \QA \to \QA^\prime.\] Since $\(Z, F_Z\)$ is the right foot of $\(S, F_S\)$ and the left foot of $\(Q, F_Q\)$, \[\(\SR^\prime, F_{\SR^\prime}\) = \(\QL^\prime, F_{\QL^\prime}\) = \(Z, F_Z\).\] If $P$ is an $\F$-pullback of $\(\sR^\prime, \qL^\prime\)$, then there is a span isomorphism $\Phi$ with \[\Phi \colon X\times_ZY \to \PA.\]  The augmented span $\(S^\prime,F_{S^\prime}\)\circ_P \(Q^\prime, F_{Q^\prime}\)$ is given by %
\begin{align*}\(S^\prime,F_{S^\prime}\)\circ_P \(Q^\prime, F_{Q^\prime}\) &= \(\(\sL^\prime\circ\pL, \qR^\prime\circ\pR\), F_{S^\prime\circ_P Q^\prime}\),\end{align*} where \[F_{S^\prime\circ_P Q^\prime} = \(F_{\SA^\prime}\circ\pL+ F_{\QA^\prime}\circ\pR - F_Z\circ{\sR^\prime}\circ\pL, F_V, F_W\).\]  Since $\alpha$ and $\beta$ are isomorphisms of augmented spans, \[F_{\SA^\prime}\circ \alpha  = F_X \quad {\rm and}\quad F_{\QA^\prime}\circ \beta  = F_Y.\] The function $\Phi$ is a span isomorphism and so \[\pL\circ \Phi = \alpha\circ \pi_{X} \quad {\rm and} \quad \pR\circ \Phi = \beta\circ \pi_Y,\] hence \[F_{\SA^\prime} \circ \pL\circ \Phi = F_{\SA^\prime}\circ \alpha\circ\pi_X = F_X \circ\pi_{X}.\] Similar arguments show that  \[F_{\QA^\prime} \circ \pR\circ \Phi = F_Y \circ\pi_{Y} \quad {\rm and}\quad F_Z\circ{\sR^\prime}\circ\pL \circ \Phi= F_{Z}\circ \pi_Z,\] and so \begin{equation}\label{eq:FSQFSPQ}F_{S\circ Q} = \(F_{S^\prime\circ_P Q^\prime}\)\circ \Phi.\end{equation} Equality \eqref{eq:FSQFSPQ} implies that $\Phi$ is an augmented span isomorphism, hence the composition of $\left[S,F_{S}\right]$ and $\left[Q,F_{Q}\right]$ is independent of representative.  The composite $\left[S,F_{S}\right] \circ \left[Q,F_{Q}\right]$ is, therefore, a well defined morphism from $(\QR, F_{\QR})$ to $(\SL, F_{\SL})$.

(ii) Let $\left[S, F_S\right]$ be a morphism with source $\(\SR, F_{\SR}\)$ and target $\(\SL, F_{\SL}\)$.  Let $\big({\rm I}_{\SR}, F_{{\rm I}_{\SR}}\big)$ be a representative of the identity augmented span with source $\(\SR, F_{\SR}\)$ and target $\(\SR, F_{\SR}\)$.  The equality \[[S]\circ[{\rm I}_{\SR}] = [S]\] follows from the fact that both ${\rm Span}\!\(\SympSurj, \F\)$ and ${\rm Span}\!\(\Riem, \F\)$ are categories.  Let the span $P$ be an $\F$-pullback of $\(\sR, {\rm Id}_{\SR}\)$, where \[\PL = \PA = \SA, \; \PR = \SR,\; \pL = {\rm Id}_{\SA},\;{\rm and}\; \pR = \sR.\]  The equalities \begin{align*}F_{\PA} &= F_{\SA}\circ\pL + F_{\SR}\circ\sR - F_{\SR}\circ{\sR}\circ{\pL} \\& = F_{\SA}\circ{\rm Id_{\SA}} + F_{\SR}\circ\sR - F_{\SR}\circ{\sR}\circ{\rm Id}_{\SA} = F_{\SA}\end{align*}  imply that there is an augmented span isomorphism from $\(S, F_S\)\circ \({\rm I}_{\SR}, F_{\SR}\)$ to $\(S, F_S\)$, and so \[\left[S, F_S\right]\circ \left[{\rm I}_{\SR}, F_{\SR}\right] = \left[S, F_S\right].\]  A similar argument shows that \[\left[{\rm I}_{\SL}, F_{\SL}\right]\circ\left[S, F_S\right] = \left[S, F_S\right].\] Therefore, $\HamSy$ and $\LagSy$ have left and right unit laws.

(iii) Refer to the following diagram for the naming of the maps below, where all spans paired with a given cospan are augmented $\F$-pullbacks of the given cospan and the diagram is commutative:
\begin{center}
\begin{showTikz}
\begin{tikzpicture}[->=stealth',node distance=2.25cm, auto]
\node(SL) at (-1.5,0) {$\SL$};
\node(SR) at (1.5,0) {$\SR = \QL$};
\node(SA) at (0,1.75) {$\SA$};
\draw[->] (SA) to node[inner sep = 8, above] {$\sL$} (SL);
\draw[->] (SA) to node[inner sep = 8, above] {$\sR$} (SR);

\node(QR) at (4.5,0) {$\QR = \TL$};
\node(QA) at (3,1.75) {$\QA$};
\draw[->] (QA) to node[inner sep = 8, above] {$\qL$} (SR);
\draw[->] (QA) to node[inner sep = 8, above] {$\qR$} (QR);

\node(TR) at (7.5,0) {$\TR$};
\node(TA) at (6,1.75) {$\TA$};
\draw[->] (TA) to node[inner sep = 8, above] {$\tL$} (QR);
\draw[->] (TA) to node[inner sep = 8, above] {$\tR$} (TR);

\node(PAa) at (1.5,3.5) {$\PA^{\1}$};
\draw[->] (PAa) to node[inner sep = 8, above] {$\pL^{\1}$\mbox{\;}} (SA);
\draw[->] (PAa) to node[inner sep = 8, above] {\mbox{\;\;\;\;\;}$\pR^{\1}$} (QA);

\node(PAb) at (4.5,3.5) {$\PA^{\2}$};
\draw[->] (PAb) to node[inner sep = 8, above] {$\pL^{\2}$\mbox{\;}} (QA);
\draw[->] (PAb) to node[inner sep = 8, above] {\mbox{\;\;\;\;\;}$\pR^{\2}$} (TA);

\node(PA) at (3,5.25) {$\PA^{\3}$};
\draw[->] (PA) to node[inner sep = 8, above] {$\pL^{\3}$} (PAa);
\draw[->] (PA) to node[inner sep = 8, above] {\mbox{\;\;\;\;\;}$\pR^{\3}$} (PAb);

\node(PAc) at (6.75,5.25) {$\PA^{\4}$};
\node[] at (5, 6.35) {$\pL^{\4}$};
\node[] at (7.35, 3.1) {$\pR^{\4}$};
\draw[->] (PAc) to node[inner sep = 8, above] {$\Phi$} (PA);

\draw[] (PAc) edge[out=-75,in=45, looseness = 1.1, dashed] (TA);
\draw[] (PAc) edge[out=150,in=100, looseness = 1.1, dashed] (PAa);
\draw[] (PAc) edge[out=225,in=85, looseness = 1.1, dashed] (QR);

\node[inner sep = 2, fill = white] at (5.85, 4.3) {$m^{\4}$};

\draw[->] (PAa) -- (SR);
\node[inner sep = 2, fill = white] (ma) at ($(PAa)!0.35!(SR)$) {$m^{\1}$};

\draw[->] (PA) -- (QA);
\node[inner sep = 2, fill = white] (ma) at ($(PA)!0.35!(QA)$) {$m^{\3}$};

\draw[->] (PAb) -- (QR);
\node[inner sep = 2, fill = white] (ma) at ($(PAb)!0.35!(QR)$) {$m^{\2}$};

\end{tikzpicture}
\end{showTikz}
\begin{showPDF}
\includegraphics{Figure8.pdf}
\end{showPDF}
\end{center}
%
%
%
Let $(P^{\3},F_{P^{\3}})$ be an $\F$-pullback of $(\pR^{\1}, \pL^{\2})$ and let $(P^{\4}, F_{P^{\4}})$ be an $\F$-pullback of $(\qR\circ\pR^{\1}, \tL)$.  

To prove (iii), show first that there is an augmented span isomorphism from the augmented span $\big((S, F_S)\circ_{(P^{\1},F_{P^{\1}})} (Q, F_Q)\big)\circ_{(P^{\4},F_{P^{\4}})}(T, F_T)$ to the augmented span $(P, F_P)$ that is given by the composite $\big((S,F_S)\circ_{(P^{\1},F_{P^{\1}})}(Q,F_Q)\big)\circ_{(P^{\3},F_{P^{\3}})}\big((Q,F_Q)\circ_{(P^{\2},F_{P^{\2}})}(T,F_T)\big)$.  A similar argument will show that there is an augmented span isomorphism from the augmented span $\(S, F_S\)\circ \big((Q, F_Q)\circ(T, F_T)\big)$ to $(P,F_P)$ and the result follows by the fact that inverses and compositions of augmented span isomorphisms are augmented span isomorphisms.  Since Lemma~5.3 of \cite{WY} proves the existence of a span isomorphism between the non-augmented spans, it suffices to show that this span isomorphism is compatible with the augmentations for the two composite spans.

The commutativity of the diagram above and the definition of the composition of augmented spans together imply that %
\begin{align*}F_{\PA^{\4}} &= F_{\PA^{\1}} \circ \pL^{\4} + F_{\TA} \circ \pR^{\4} - F_{\QR} \circ m^{\4} 
\\&= F_{\PA^{\1}} \circ \pL^{\3}\circ \Phi + F_{\TA} \circ\pR^{\2}\circ \pR^{\3}\circ \Phi - F_{\QR} \circ m^{\2} \circ\pR^{\3}\circ\Phi.
\\&= \(F_{\PA^{\1}} \circ \pL^{\3} + F_{\TA} \circ\pR^{\2}\circ \pR^{\3} - F_{\QR} \circ m^{\2} \circ\pR^{\3}\)\circ \Phi
\\&= \(F_{\PA^{\1}} \circ \pL^{\3} + \(F_{\TA} \circ\pR^{\2} - F_{\QR} \circ m^{\2}\) \circ\pR^{\3}\)\circ \Phi
\\&= \(F_{\PA^{\1}} \circ \pL^{\3} + \(F_{\QA} \circ\pL^{\2} -F_{\QA} \circ\pL^{\2} +F_{\TA} \circ\pR^{\2} - F_{\QR} \circ m^{\2}\) \circ\pR^{\3}\)\circ \Phi
\\& = \(F_{\PA^{\1}} \circ \pL^{\3} + \(F_{\QA} \circ\pL^{\2} +F_{\TA} \circ\pR^{\2} - F_{\QR} \circ m^{\2}\) \circ\pR^{\3} -F_{\QA} \circ\pL^{\2}\circ \pR^{\3}\)\circ \Phi
\\&= \(F_{\PA^{\1}} \circ \pL^{\3} + \(F_{\QA} \circ\pL^{\2} +F_{\TA} \circ\pR^{\2} - F_{\QR} \circ m^{\2}\) \circ\pR^{\3}  -F_{\QA} \circ m^{\3}\)\circ \Phi
\\&= \(F_{\PA^{\1}} \circ \pL^{\3} + F_{P^{\2}} \circ\pR^{\3}  -F_{\QA} \circ m^{\3}\)\circ \Phi
= F_{\PA^{\3}}\circ \Phi.
\end{align*} 
Therefore, the span isomorphism $\Phi$ is compatible with the augmentations $F_{P^{\4}}$ and $F_{P^{\3}}$.
{}{}{}{}{}{}{}
\end{proof}

\subsection{Motivating Example}

Suppose that the spring-mass system with three masses given in Section~\ref{sec:intro} has masses $m_1$, $m_2$, and $m_3$ respectively as the left, middle, and right masses  of the system.  Suppose further that the spring constants of the left and right springs are respectively $k_1$ and $k_2$.  The spring-mass system with three masses is a composite of two spring-mass systems with two masses each.  We now discuss a category theoretic construction of a model for the composite system with its subsystems.

Let $[S, V_S]$ be a Lagrangian system describing the left-spring mass system and $[Q, V_Q]$ be a Lagrangian systems describing the right spring-mass system.  Denote both $\SR$ and $\QL$ by $Z$, since $\SR$ is equal to $\QL$, and by $V_Z$ the augmentation on $Z$.  Take a representative $(S, V_S)$ of the Langrangian system $[S, V_S]$ to be the Riemannian span with the manifold $\SA$ equal to $\mathbb R^2$ and the manifolds $\SL$ and $Z$ equal to $\mathbb R$.  Let $g_1$ be the standard Riemannian metric on $\mathbb R$.  Let $\rho_L$ and $\rho_R$ be the canonical projections on $\mathbb R^2$ with \[\rho_L(q_1, q_2) = q_1 \quad {\rm and}\quad \rho_R(q_1, q_2) = q_2.\]  Denote by $g_2$ the standard Riemannian metric on $\mathbb R^2$.  Endow $\SL$ with the Riemannian metric $g_{\SL}$ and $Z$ with the Riemannian metric $g_Z$, where $g_{\SL}$ and $g_Z$ are given by \[g_{\SL} = m_1g_1 \quad {\rm and} \quad g_Z = m_2g_1.\]  Define by $g_{\SA}$ the metric on $\mathbb R^2$ given for all $v$ and $w$ in $T_{(q_1,q_2)}\mathbb R^2$ by \[g_{\SA}(v,w) = g_{\SL}({\rm d}\rho_L(v), {\rm d}\rho_L(w)) + g_Z({\rm d}\rho_R(v), {\rm d}\rho_R(w)).\] Denote respectively by $\sL$ and $\sR$ the functions from $\SA$ to $\SL$ and from $\SA$ to $Z$ that act on underlying manifolds as the projections $\rho_L$ and $\rho_R$.  The augmentation $V_S$ is the triple of maps \[V_S = (V_{\SA}, V_{\SL}, V_Z) \quad {\rm with}\quad V_{\SA}(q_1,q_2) = \frac{k_1}{2}(q_1-q_2)^2,\; V_{\SL} \equiv 0,\; {\rm and}\; V_{Z} \equiv 0.\]  Define similarly the Riemannian span $(Q, V_Q)$, but with the Riemannian metric $g_{\QR}$ on $\QR$ and the augmentations $V_{\QA}$ and $V_{\QR}$ given by \[g_{\QR} = m_3g_{1},\; V_{\QA}(q_2,q_3) = \frac{k_2}{2}(q_2-q_3)^2, \; {\rm and}\; V_{\QR} \equiv 0.\]    Define by $g_{\QA}$ the metric on $\mathbb R^2$ given for all $v$ and $w$ in $T_{(q_2,q_3)}\mathbb R^2$ by \[g_{\QA}(v,w) = g_{Z}({\rm d}\rho_L(v), {\rm d}\rho_L(w)) + g_{\QR}({\rm d}\rho_R(v), {\rm d}\rho_R(w)).\]

Denote by $\pi_L$ and $\pi_R$ the respective projections from $\SA\times_Z\QA$ to $\SA$ and to $\QA$ and by $\pi_M$ the map $\sR \circ \pi_L$, which is also the map $\qR\circ \pi_R$.   Denote by $g_{\SA\times_Z\QA}$ the Riemannian metric on $\SA\times_Z\QA$ given by \[g_{\SA\times_Z\QA} = \pi_L^\ast(g_{\SA}) + \pi_R^\ast(g_{\QA}) - \pi_M^\ast(g_Z).\]  The augmentation $V_{\SA\times_Z\QA}$ is then given by \[V_{\SA\times_Z\QA} = \pi_L^\ast(V_{\SA}) + \pi_R^\ast(V_{\QA}) - \pi_M^\ast(V_{Z}).\]  Let $\Phi$ be the diffeomorphism from $\SA\times_Z\QA$ to $\mathbb R^3$ given by \[\Phi(q_1,q_2,q_2,q_3)=(q_1,q_2,q_3),\] so that ${\rm d}\Phi$ is the diffeomorphism from $T(\SA\times_Z\QA)$ to $T\mathbb R^3$ given by\[{\rm d}\Phi(q_1,q_2,q_2,q_3, \dot{q}_1,\dot{q}_2,\dot{q}_2,\dot{q}_3)=(q_1,q_2,q_3, \dot{q}_1,\dot{q}_2,\dot{q}_3).\]     The configuration space of the composite system is, up to an isometry, the fibered product of the configuration spaces of the open subsystems:

\begin{center}
\begin{showTikz}
\begin{tikzpicture}

\coordinate(r1) at (-1.25,2.5);
\coordinate(r2) at (0,2.5);
\coordinate(r3) at (1.25,2.5);

\coordinate(r4) at (-1.25,1.25);
\coordinate(ra) at (-2.25,1.25);
\coordinate(rb) at (-.75,1.25);

\coordinate(r5) at (1.25,1.25);
\coordinate(rc) at (.75,1.5);
\coordinate(rd) at (2.25,1.5);

\coordinate(br) at (-2.5,0);
\coordinate(bm) at (0,0);
\coordinate(bl) at (2.5,0);

\coordinate(l1) at (-.75,2.25);
\coordinate(l2) at (.75,3);
\coordinate(l3) at (-1.25,1.25);
\coordinate(l4) at (1.25,1.25);

\node[] at (r2) {${\mathbb R^3}$};
\node[] at (l3) {${\mathbb R^2}$};
\node[] at (l4) {$ {\mathbb R^2}$};
\node[] at (bl) {$ {\mathbb R}$};
\node[] at (bm) {$ {\mathbb R}$};
\node[] at (br) {${\mathbb R}$};

{\draw[shorten >=10pt,shorten <=15pt, ->, arrowhead=5pt, line width=.5pt] (r2) -- (r4);}
{\draw[shorten >=10pt,shorten <=15pt, ->, arrowhead=5pt, line width=.5pt] (r2) -- (r5);}
{\draw[shorten >=10pt,shorten <=15pt, ->, arrowhead=5pt, line width=.5pt] (r4) -- (br);}
{\draw[shorten >=10pt,shorten <=15pt, ->, arrowhead=5pt, line width=.5pt] (r4) -- (bm);}
{\draw[shorten >=10pt,shorten <=15pt, ->, arrowhead=5pt, line width=.5pt] (r5) -- (bm);}
{\draw[shorten >=10pt,shorten <=15pt, ->, arrowhead=5pt, line width=.5pt] (r5) -- (bl);}
\end{tikzpicture}
\end{showTikz}
\begin{showPDF}
\includegraphics{Figure9.pdf}
\end{showPDF}
\end{center}
%
%
%

Denote by $\PA$ the Riemannian manifold $\mathbb R^3$, and by $\pL$ and $\pR$ the maps \[\pL = \sL\circ \pi_L\circ \Phi^{-1} \quad {\rm and}\quad \pR = \sR\circ \pi_R\circ \Phi^{-1}.\]
Denote by $V_{\PA}$ the potential \[V_{\PA} = V_{\SA\times_Z\QA}\circ \Phi^{-1}.\] Define a Riemannian metric $g_{\PA}$ on $\PA$ by \[g_{\PA} = (\Phi^{-1})^\ast(g_{\SA\times_Z\QA}),\] making $\Phi$ an isometry. The Lagrangian for the composite system is ${\mathcal L}_{\PA}$ where for every $\nu$ in $T\PA$, 
 \[{\mathcal L}_{\PA}(\nu) = \frac{1}{2}g_{\PA}\!\(\nu, \nu\) - V_{\PA}\!\(\proj{\PA}\!\(\nu\)\).\]  The Lagrangian ${\mathcal L}$ of the system with configuration space given by $\mathbb R^3$ is given with respect to coordinate system $(q_1,q_2,q_3)$ by%
 \begin{align*}
 {\mathcal L}(q_1,q_2,q_3, \dot{q}_1,\dot{q}_2,\dot{q}_3) &= \frac{m_1}{2}(\dot{q}_1)^2 + \frac{m_2}{2}(\dot{q}_2)^2 + \frac{m_2}{2}(\dot{q}_2)^2 +\frac{m_3}{2}(\dot{q}_3)^2 - \frac{m_2}{2}(\dot{q}_2)^2\\&\qquad  - \frac{k_1}{2}(q_1 - q_2)^2 - \frac{k_1}{2}(q_2 - q_3)^2 + 0 \quad (\text{since }V_Z \equiv 0)\\& = \frac{m_1}{2}(\dot{q}_1)^2 + \frac{m_2}{2}(\dot{q}_2)^2  +\frac{m_3}{2}(\dot{q}_3)^2  - \frac{k_1}{2}(q_1 - q_2)^2 - \frac{k_1}{2}(q_2 - q_3)^2.
 \end{align*}  The Riemannian span $(P, V_P)$ is a representative of the Lagrangian system $[S,V_S]\circ[Q, V_Q]$.  The Lagrangian $\mathcal L$ on $\PA$ is the Lagrangian for the given system of three masses and two springs with configuration space equal to $\mathbb R^3$.   We leave the determination of the Hamiltonian system to the reader as it is a straightforward exercise given the previous discussion and the result of the next section.
 
In general, a description of a composite system requires a prior description of the subsystems.  The subsystems need not themselves have descriptions as composite systems and it remains an open problem to determine the simplest subsystems that are required to construct from them any other system as a composite. If two subsystems that share a common component form a complicated system, and if we know how to map the subsystems into two pieces, one of which is the common component, then we can view the complicated system as a composite system in our formalism.  We systematically work through a selection of examples in an upcoming paper where we more carefully develop computational tools. 
 
\section{The Legendre Functor}\label{sec:Functor}

This section constructs a functor $\FLag$ from $\LagSy$ to $\HamSy$ that preserves the paths of motion.

Suppose that $\(M, g_M\)$ is a Riemannian manifold of dimension $m$.  The canonical 2-form, $\omega_{T^\ast M}$, is the exterior derivative of the tautological 1-form and is a symplectic form on $T^\ast M$, \cite[p. 202]{Arn}.  Denote respectively by $\dualproj{M}$ and $\proj{M}$ the canonical projections from $T^\ast M$ to $M$ and from $TM$ to $M$. Suppose $a$ is a point of $M$.  There is an open set $U$ of $M$ containing $a$ that is the domain of coordinates $\(x_i\)_{i\in\{1,\dots, m\}}$.   The set of 1-forms $\{{\rm d}x_i\colon i\in \{1, \dots, m\}\}$ trivializes the subbundle $T^\ast U$.  Define for each $i$ the real valued functions $p^M_i$ on $T^\ast U$ with the property that for all $\theta$ in $T^\ast M$, \[\theta = \sum_{i=1}^m \left.p^M_i(\theta)\,{\rm d} x_i\right|_{\dualproj{M}(\theta)}.\] The $p^M_i$ are the \emph{momenta} associated with the $x_i$ coordinates.  For each $i$, the function $p^M_i$ is the evaluation map ${\rm ev}\!\!_{\left.\frac{\partial}{\partial x_i}\right|_{\dualproj{M}(\theta)}}$ that is defined by the equality \[{\rm ev}\!\!_{\left.\frac{\partial}{\partial x_i}\right|_{\pi_M(\theta)}}(\theta) = \theta\(\left.\frac{\partial}{\partial x_i}\right|_{\dualproj{M}(\theta)}\).\]  For each $i$, define $q^M_i$ by \[q_i^M = x_i\circ \pi_M.\]  The function given by $\(q_i^M, p_i^M\)_{i\in\{1, \dots, m\}}$ on $\dualproj{M}^{-1}\(U\)$ is a Darboux coordinate system, that is \[\omega_{T^\ast M} = \sum_{i=1}^m{\rm d}q^M_i\wedge{\rm d}p^M_i.\] Define for each $i$ the real valued function $\hat{q}^M_i$ on $TM$ with the property that if $v$ is in $\proj{M}^{-1}(U)$, then \[v = \sum_{i=1}^m \hat{q}^M_i(v)\left.\frac{\partial}{\partial x_i}\right|_{\proj{M}(v)}.\]  Note that $\hat{q}^M_i$ is the function defined for each $v$ in $TU$ by \[\hat{q}^M_i(v) = \left.{\rm d}x_i\right|_{\proj{M}(v)}(v).\] Denote ambiguously by $q_i^M$ the function \[q_i^M = x_i\circ\proj{M}\] on $TU$.  The coordinate system $\(q_i^M, \hat{q}^M_i\)$ is a coordinate system on $\proj{M}^{-1}\!\(\pi_M\!\(U\)\)$.

The Riemannian metric $\met{M}$ on $TM$ induces a Riemannian metric on the cotangent bundle $T^\ast M$, to be denoted $\dualmet{M}$ and for each $a$ in $U$ defined on the pair $(\theta_1, \theta_2)$ in $T^\ast_aM\times T^\ast_aM$ by \begin{align*}\dualmet{M}(\theta_1, \theta_2) = \met{M}(\ginv_M(\theta_1), \ginv_M(\theta_2)) = \sum_{i,j = 1}^m\met{M}^{ij}(a)p_i^M(\theta_1)p_j^M(\theta_2),\end{align*} where $\met{M}^{ij}$ denotes the $(i,j)$ entry of the inverse of the matrix given by $\met{M}$ in the $(q_i^M, \hat{q}^M_i)$ coordinates.  For all $v$ in $TM$ and $\theta$ in $T^\ast M$, denote respectively by $\met{M}(\cdot)$ and $\dualmet{M}(\cdot)$ the quadratic forms \begin{equation}\label{eq:qform}\met{M}(v) = \met{M}(v,v)\quad {\rm and}\quad \dualmet{M}(\theta) = \dualmet{M}(\theta,\theta).\end{equation}

Define $\preFLag$ as a map from Riemannian manifolds to symplectic manifolds by \[\preFLag\!\(M, g_M\) = \(T^\ast M, \omega_{T^\ast M}\).\]  For any surjective Riemannian submersion $f$ from $M$ to $N$, define $\preFLag\!\(f\)$ by \[\preFLag\!\(f\) = \flat_N\circ {\rm d}f\circ\sharp_M.\]  To simplify the notation, denote by $F$ the function $\preFLag\!\(f\)$.  We depict the various maps here:

 \begin{center}
 \begin{showTikz}
\begin{tikzpicture}[node distance=2.25cm, auto]

\node(W) at (0,0) {$T^{*}M$};
\node(Wr) at (.25,0) {\phantom{$T^{*}M$}};
\node(X) at (0,2) {$TM$};
\node(Xr) at (.25,2) {\phantom{$TM$}};

\node(Y) at (4,2) {$TN$};
\node(Z) at (4,0) {$T^{*}N$};

\node(Yl) at (3.75,2) {\phantom{$TN$}};
\node(Zl) at (3.75,0) {\phantom{$T^{*}N$}};

\node(A) at (0,4) {$M$};
\node(B) at (4,4) {$N$};
\draw[->] (W) to node[left]{$\sharp_{M}$}(X);
\draw[->] (Y)to node[right] {$\flat_{N}$}(Z);

\draw[->, dashed] (Xr) to node[right]{$\flat_{M}$}(Wr);
\draw[->, dashed] (Zl) to node[left] {$\sharp_{N}$}(Yl);

\draw[->] (W)to node[below] {$F = \preFLag(f)$}(Z);
\draw[->] (A) to node[above] {$f$}(B);
\draw[->] (X) to node[above] {${\rm d}f$}(Y);
\draw[->] (X) to node[left] {$\proj{M}$}(A);
\draw[->] (Y) to node[right] {$\proj{N}$}(B);

\draw[->] (W) edge[out=150,in=210, looseness = 1.1] node[left] {$\dualproj{M}$}(A);
\draw[->] (Z) edge[out=30,in=330, looseness = 1.1] node[right] {$\dualproj{N}$}(B);

\end{tikzpicture}
\end{showTikz}
\begin{showPDF}
\includegraphics{Figure10.pdf}
\end{showPDF}
\end{center}
%

Suppose that $M$ and $N$ are Riemannian manifolds of respective dimensions $m$ and $n$ and suppose further that $f$ is a surjective Riemannian submersion from $M$ to $N$.  For any point $p$ in $M$ there is a coordinate system $(x_1, \dots, x_m)$ of $M$ on an open set containing $p$ and a coordinate system $(y_1, \dots, y_n)$ of $N$ on an open set containing $f(p)$ such that for all $i$ in $\{1, \dots, n\}$ and $k$ in $\{n+1, \dots, m\}$, \[x_i = y_i\circ f\quad \text{and}\quad \frac{\partial}{\partial x_k}\in {\rm ker}({\rm d}f).\]  Let $j$ be an index varying in the set $\{1, \dots, n\}$. For each $i$ and each $j$, denote respectively by $q^M_i$ and $q^N_j$ the functions $x_i\circ \dualproj{M}$ and $y_j\circ\dualproj{N}$ and denote by $p^M_i$ and $p^N_j$ the momenta associated with the coordinate functions $x_i$ and $y_j$.  Use the above notation for the following lemma, as well as for the rest of the section.

\begin{lemma}\label{lem:pmpn}
For all $p^M_j$, $p^N_j$, and $F$ defined as above,  \[p^M_j = p^N_j\circ F.\]
\end{lemma}
 
\begin{proof}
For all $j$ in $\{1, \dots, n\}$, \[{\rm d}f\(\left.\frac{\partial}{\partial x_j}\right|_a\)  = {\rm d}f\(\left.\frac{\partial}{\partial (y_j \circ f)}\right|_a\) = \left.\frac{\partial}{\partial y_j}\right|_{f(a)}.\]

For all $\theta$ in $T^\ast U$, there is an element $X$ of $TU$ with $\theta$ equal to $g_M(X, \cdot)$.  In this case, the form $F(\theta)$ is equal to $g_N({\rm d}f(X), \cdot)$, and so 
\begin{align*}
p^M_j(\theta) = {\rm ev}\!\!_{\left.\frac{\partial}{\partial x_j}\right|_{\dualproj{M}(\theta)}}(\theta)= g_M\(X, \left.\frac{\partial}{\partial x_j}\right|_{\dualproj{M}(\theta)}\).
\end{align*}  The function $f$ is a surjective Riemannian submersion, implying that
\begin{equation*}
g_M\(X, \left.\frac{\partial}{\partial (y_j\circ f)}\right|_{\dualproj{M}(\theta)}\) = g_N\({\rm d}f(X), {\rm d}f\!\(\left.\frac{\partial}{\partial (y_j\circ f)}\right|_{\dualproj{M}(\theta)}\)\)
\end{equation*} and so 
\begin{align*}
p^M_j(\theta) &= g_N\({\rm d}f(X), \left.\frac{\partial}{\partial y_j}\right|_{f(\dualproj{M}(\theta))}\)\\
& = g_N\({\rm d}f(X), \left.\frac{\partial}{\partial y_j}\right|_{\dualproj{N}(F(\theta))}\)\\
& = F(\theta)\left(\left.\frac{\partial}{\partial y_j}\right|_{\dualproj{N}(F(\theta))}\right)\\
& = {\rm ev}\!\!_{\left.\frac{\partial}{\partial y_j}\right|_{\dualproj{N}(F(\theta))}}(F(\theta)) = (p^N_j\circ F)(\theta),
\end{align*}
which proves the desired equality.
\end{proof}

\begin{proposition}\label{prop:FLagfromRiemtoPoissonMaps}
For any surjective Riemannian submersion $f$ from a Riemannian manifold $M$ to a Riemannian manifold $N$, the function $\preFLag(f)$ is a surjective Poisson map.
\end{proposition}

\begin{proof}
Suppose $M$ and $N$ have respective dimensions $m$ and $n$.  The map $\preFLag$ maps Riemannian manifolds to symplectic manifolds, in particular, \[\preFLag(M) = T^\ast M \quad{\rm and}\quad \preFLag(N) = T^\ast N.\]   Once again denote by $F$ the map $\preFLag(f)$.  Suppose that $\Pi_{T^\ast M}$ and $\Pi_{T^\ast N}$ respectively denote the Poisson bivectors for ${T^\ast M}$ and ${T^\ast N}$. For any $\alpha$ and $\beta$ in $C^{\infty}(T^\ast N)$ and any $a$ in $T^\ast M$,
\begin{align}\label{eq:riemtopois}
{\rm d}F_{a}(\Pi_{T^{\ast}M})({\rm d}\alpha,{\rm d}\beta)&=\Pi_{T^{\ast}M}({\rm d}(\alpha\circ F), {\rm d}(\beta\circ F)){\Big|}_{a}\notag\\
&=\sum_{i=1}^m\left.\left(\frac{\partial{(\alpha\circ F)}}{\partial{q^{M}_{i}}}\frac{\partial{(\beta\circ F)}}{\partial{p^{M}_{i}}}-\frac{\partial{(\beta\circ F)}}{\partial{q^{M}_{i}}}\frac{\partial{(\alpha\circ F)}}{\partial{p^{M}_{i}}}\right)\right|_{a}
\notag\\&=\sum_{i=1}^n\left.\(\frac{\partial{(\alpha\circ F)}}{\partial{q^{M}_{i}}}\frac{\partial{(\beta\circ F)}}{\partial{p^{M}_{i}}}-\frac{\partial{(\beta\circ F)}}{\partial{q^{M}_{i}}}\frac{\partial{(\alpha\circ F)}}{\partial{p^{M}_{i}}}\)\right|_{a}\notag\\
&=\sum_{i=1}^n\left.\(\frac{\partial{(\alpha\circ F)}}{\partial{(q^{N}_{i}\circ F)}}\frac{\partial{(\beta\circ F)}}{\partial{(p^{N}_{i}\circ F)}}-\frac{\partial{(\beta\circ F)}}{\partial{(q^{N}_{i}\circ F)}}\frac{\partial{(\alpha\circ F)}}{\partial{(p^{N}_{i}\circ F)}}\)\right|_{a}\\
&=\sum_{i=1}^n\left.\(\frac{\partial{\alpha}}{\partial{q^{N}_{i}}}\frac{\partial{\beta}}{\partial{p^{N}_{i}}}-\frac{\partial{\beta}}{\partial{q^{N}_{i}}}\frac{\partial{\alpha}}{\partial{p^{N}_{i}}}\)\right|_{F(a)} =\Pi_{T^{\ast}N}({\rm d}\alpha, {\rm d}\beta){\Big|}_{F(a)},
\notag\end{align} where Lemma~\ref{lem:pmpn} implies the equality in \eqref{eq:riemtopois}. Therefore, ${\rm d}F(\Pi_{T^{\ast}M})$ is equal to $\Pi_{T^{\ast}N}$, which implies that $F$ is a Poisson map.  The map $f$ is a surjective submersion, therefore ${\rm d}f$ is surjective.  The nondegeneracy of $g$ implies that $F$ is also surjective and so $\preFLag$ maps the morphisms in $\Riem$ to morphisms in $\SympSurj$.
\end{proof}

\begin{lemma}\label{lem:FlagofAug}
For any spans $S$ and $Q$ in $\Riem$ and any span isomorphism $\Phi$ from $S$ to $Q$, the function $\preFLag(\Phi)$ is a span isomorphism from $\preFLag(S)$ to $\preFLag(Q)$.
\end{lemma}

\begin{proof}
Suppose that $\Phi$ is a span isomorphism from $S$ and $Q$.  In this case, $\preFLag(\Phi)$ is Poisson.  Since $\preFLag(\Phi)$ is a diffeomorphism and Poisson, it is an isomorphism in the category $\SympSurj$.  Recall that the isomorphisms in $\SympSurj$ are Poisson diffeomorphisms, which are symplectomorphisms since the objects in $\SympSurj$ are symplectic manifolds, \cite[p. 195]{AM}.  Since $\Phi$ is a span morphism, \[\sL = \qL\circ \Phi \quad {\rm and}\quad \sR = \qR\circ \Phi,\] implying that %
\begin{align*}\preFLag(\sL) &= \preFLag(\qL\circ \Phi) 
\\& =\flat_{\QL}\circ{\rm d}\!\(\qL\circ \Phi\)\circ\sharp_{\SA}
\\& =\flat_{\QL}\circ{\rm d}\qL\circ {\rm d}\Phi\circ\sharp_{\SA}
\\& =\flat_{\QL}\circ{\rm d}\qL\circ\(\sharp_{\QA}\circ\flat_{\QA}\)\circ {\rm d}\Phi\circ\sharp_{\SA}
\\& =\(\flat_{\QL}\circ{\rm d}\qL\circ\sharp_{\QA}\)\circ\(\flat_{\QA}\circ {\rm d}\Phi\circ\sharp_{\SA}\) = \preFLag(\qL)\circ \preFLag(\Phi).
\end{align*} %
A similar argument shows that \[\preFLag(\sR) = \preFLag(\qR)\circ \preFLag(\Phi),\] proving that $\preFLag(\Phi)$ is a span morphism.  Therefore, for any spans $S$ and $Q$ in $\Riem$ that are span isomorphic, the spans $\preFLag(S)$ and $\preFLag(Q)$ are also span isomorphic.  
\end{proof}

\begin{lemma}\label{lem:FlagofAug}
Suppose that $(S, V_S)$ and $(Q, V_Q)$ are Riemannian spans and that $\Phi$ is an isomorphism of spans from $S$ to $Q$.  If $\Phi$ is additionally an isomorphism of classical spans, then so is $\preFLag(\Phi)$.
\end{lemma}

\begin{proof}
In light of Lemma~\ref{lem:FlagofAug}, it suffices to show that $\preFLag(\Phi)$ is compatible with the augmentations.  For any span isomorphism $\Phi$ from $S$ to $Q$ that is compatible with $V_S$ and $V_Q$, \[V_{\SA} = V_{\QA}\circ \Phi.\] The isomorphism $\Phi$ is Riemannian, hence an isometry.  Therefore, \[\dualmet{\SA} = \dualmet{\QA}\circ \preFLag(\Phi),\] and so
\begin{align*}
H_{\SA} &= \frac{1}{2}\dualmet{\SA} + V_{\SA}\circ\pi_{\SA}\\
& =  \frac{1}{2}\dualmet{\QA}\circ \preFLag(\Phi) + V_{\QA}\circ \pi_{\QA}\circ \preFLag(\Phi) = H_{\QA} \circ \preFLag(\Phi).
\end{align*}
\end{proof}

Suppose that $(S, V_S)$ is a Riemannian span and let $\star$ denote either of the letters $A$, $L$, or $R$.   Define $\preFLag\!\(S_\star, V_{S_\star}\)$ by \[\preFLag\!\(S_\star, V_{S_\star}\) = \(\preFLag\!\(S_\star\), H_{S_\star}\)\] where for all $\eta$ in $S_\star$, \[H_{S_\star}\!(\eta) = \frac{1}{2}\dualmet{S_\star}(\eta) + (V_{S_\star}\circ \dualproj{S_\star})(\eta).\] Each object of $\LagSy$ is an augmented Riemannian manifold and so $\preFLag$ maps the objects of $\LagSy$ to the objects of $\HamSy$, and the morphisms of $\Riem$ to the morphisms of $\SympSurj$.  In this way, $\preFLag$ maps Riemannian spans to Poisson spans.  Define $\FLag$ to be $\preFLag$ on the objects of $\LagSy$ and for each morphism $[S,V_S]$ in $\LagSy$, define $\FLag([S, V_S])$ by \[\FLag([S, V_S]) = [\preFLag(S, V_S)].\] 

\begin{theorem}
The map $\FLag$ is a functor from $\LagSy$ to $\HamSy$.  Suppose that $\pi_{\SA}$ is the canonical projection from $T^\ast\SA$ to $\SA$.  Suppose that the Lagrangian system $[S,V_S]$ has a path of motion $\gamma$ on the manifold $\SA$ that is specified by the representative $(S, V_S)$ of $[S, V_S]$ and suppose that $\gamma$ intersects a point $x$ of $\SA$ at time zero.  In this case, the path $\preFLag\circ \gamma$ is a path determined by $\FLag([S, V_S])$, valued in the symplectic manifold $\preFLag(\SA)$, and $\pi_{\SA}\circ\preFLag\circ\gamma$ also intersects $x$ at time zero.
\end{theorem}

\begin{proof}
{}{}{}{}{}{}

The map $\FLag$ maps Riemannian manifolds to symplectic manifolds and potentials to Hamiltonians, and therefore maps the objects of $\LagSy$ to the objects of $\HamSy$. Proposition~\ref{prop:FLagfromRiemtoPoissonMaps} implies that $\FLag$ maps surjective Riemannian submersions to surjective Poisson maps, and so if $S$ is a span in $\Riem$, then $\preFLag(S)$ is a span in $\SympSurj$.  Lemma~\ref{lem:FlagofAug} implies that if $(S, F_S)$ and $(Q, F_Q)$ are isomorphic as Riemannian spans, then $\preFLag(S, F_S)$ and $\preFLag(Q, F_Q)$ are isomorphic as Poisson spans and so $\FLag$ is well defined on Lagrangian systems, mapping them to Hamiltonian systems.

Suppose that $M$ is a Riemannian manifold.  Denote by ${\mathcal L}_M$ the Lagrangian on $TM$, where for each $\nu$ in $TM$, \[{\mathcal L}_M(\nu) = \frac{1}{2}g_M\!\(\nu, \nu\) - V_M\!\(\proj{M}\!\(\nu\)\).\]  Denote by $H_M$ the Hamiltonian associated to $V_M$ and by $\{\cdot, \cdot\}_{T^\ast M}$ the Poisson bracket as given above in the construction of $\FLag$.  It is a standard result in classical mechanics that a path $\gamma$ on $M$ is a solution to \eqref{Equation:EL} if and only if it is an integral curve of $\{\cdot, H_M\}_M$, \cite[p.25, Theorem 3.13]{Cort}.  This proves the last two statements of the theorem. To prove that $\FLag$ is a functor, it suffices to show further that: (i) $\FLag$ preserves composition and (ii) $\FLag$ maps identity morphisms to identity morphisms.  

To show (i), suppose that $[S, F_S]$ and $[Q, F_Q]$ are Riemannian spans and that $[S, F_S]$ is composable with $[Q, F_Q]$.    Suppose that $P$ is an $\F$-pullback of $(\sR, \qL)$, where $\PA$ is the fibered product $\SA\times_{\SR}\QA$ and $\pR$ and $\pL$ are the respective restrictions of the projections on $\SA\times \QA$ to $\SA$ and $\QA$.  The map $\preFLag$ maps $\SA\times_{\SR}\QA$ to its cotangent bundle $T^\ast\!\(\SA\times_{\SR}\QA\)$, which is isomorphic in $\SympSurj$ to the manifold $(T^\ast\SA)\times_{(T^\ast\SR)}(T^\ast\QA)$. The symplectic form on $T^\ast\!\(\SA\times_{\SR}\QA\)$ is given by the canonical 2-form and the symplectic form $\omega$ on $(T^\ast\SA)\times_{(T^\ast\SR)}(T^\ast\QA)$ is given by \[\omega = \preFLag(\pL)^\ast(\omega_{T^\ast\SA}) + \preFLag(\pR)^\ast(\omega_{T^\ast\QA}) - \preFLag(\pL)^\ast(\preFLag(\sR)^\ast(\omega_{T^\ast \SR})).\]  The symplectomorphism $\Phi$ from $T^\ast\!\(\SA\times_{\SR}\QA\)$ to $(T^\ast\SA)\times_{(T^\ast\SR)}(T^\ast\QA)$ is consistent with the augmentations.  
Lemma~\ref{lem:FlagofAug} implies that 
\begin{align*}
\FLag([S, F_S]\circ[Q, F_Q]) &= \FLag([(S, F_S)\circ_P(Q, F_Q)])\\ &=  [\preFLag((S, F_S)\circ_P(Q, F_Q))] \\ &= [\preFLag(S, F_S)\circ_{\preFLag(P)}\preFLag(Q, F_Q)]\\ &= [\preFLag(S, F_S)]\circ [\preFLag(Q, F_Q)] = \FLag([S, F_S])\circ \FLag([Q, F_Q]),
\end{align*}
where the penultimate equality holds because $\preFLag(P)$ is an $\F$-pullback.

To show (ii), suppose that $(X,V_X)$ is an augmented Riemannian manifold and that ${\rm Id}_X$ is the identity map from $X$ to $X$.  Denote by ${\rm I}_X$ the span $({\rm Id}_X, {\rm Id}_X)$.  The span $\preFLag({\rm I}_X)$ is the pair $(\preFLag({\rm Id}_X), \preFLag({\rm Id}_X))$ where $\preFLag({\rm Id}_X)$ is the identity map ${\rm Id}_{T^\ast X}$ from $T^\ast X$ to $T^\ast X$.  Furthermore, $\preFLag$ maps the augmentation $V_X$ to the augmentation $H_{T^\ast X}$ where \[H_{T^\ast X} = \frac{1}{2}\dualmet{X} + V_X\circ\pi_X.\]  Suppose that $S$ is a Poisson span with $(\SL, H_{\SL})$ equal to $(T^\ast X, H_{T^\ast X})$.  Let $Q$ be the $\F$-pullback of the cospan $(\preFLag({\rm Id}_X), \sL)$ with the property that $\QA$ is the symplectic manifold $T^\ast X\times_{T^\ast X}\SA$.  The maps $\qL$ and $\qR$ are the respective restrictions to the manifold $T^\ast X\times_{T^\ast X}\SA$ of the canonical projections of the manifold $T^\ast X\times \SA$ to $T^\ast X$ and $\SA$ and are symplectomorphisms.  Since $Q$ is an $\F$-pullback, the augmentation $H_{\QA}$ on $\QA$ is given by %
\begin{align*}H_{\QA} &= \(\frac{1}{2}\dualmet{X} + V_X\circ\pi_X\)\circ \qL + \(\frac{1}{2}\dualmet{\SA} + V_{\SA}\circ\pi_{\SA}\)\circ \qR \\&\qquad- \(\frac{1}{2}\dualmet{X} + V_X\circ\pi_X\)\circ \qL\circ {\rm Id}_{T^\ast X} \\&= \(\frac{1}{2}\dualmet{X} + V_X\circ\pi_X\)\circ \qL + \(\frac{1}{2}\dualmet{\SA} + V_{\SA}\circ\pi_{\SA}\)\circ \qR - \(\frac{1}{2}\dualmet{X} + V_X\circ\pi_X\)\circ \qL\\ &=  \(\frac{1}{2}\dualmet{\SA} + V_{\SA}\circ\pi_{\SA}\)\circ \qR = H_{\SA}\circ \qR,
\end{align*}
hence \[H_{\QA} = H_{\SA}\circ \qR.\] The map $\qR$ is, therefore, compatible with the augmentations.  Since $Q$ is paired with $(\preFLag({\rm Id}_X), \sL)$, \[\sL\circ \qR = {\rm Id}_X\circ \qL = \qL,\]  and so $\qR$ is a span isomorphism mapping the composite $(\preFLag({\rm Id}_X)\circ \qL, \sR\circ \qR)$ to the span $S$ that is compatible with the augmentations. This compatibility implies that \[\FLag([{\rm I}_X, V_{{\rm I}_X}])\circ [S, H_{S}] = [\preFLag({\rm I}_X, V_{{\rm I}_X})\circ (S, H_S)] = [S, H_S].\]  Similar arguments show that for any Poisson span $(S^\prime, H_{S^\prime})$ such that $(S_R^\prime, H_{S_R^\prime})$ is equal to $(T^\ast X, H_{T^\ast X})$,  \[[S^\prime, H_{S^\prime}]\circ \FLag([{\rm I}_X, V_{{\rm I}_X}]) = [S^\prime, H_{S^\prime}],\] and so $\FLag([{\rm I}_X, V_{{\rm I}_X}])$ is the identity map with source and target $(T^\ast X, H_{T^\ast X})$.

{}{}{}{}{}{}{}
\end{proof}

We call the functor $\mathcal L$ from $\LagSy$ to $\HamSy$ the \emph{Legendre functor}.  It is a generalization of the Legendre transformation which translates from the Lagrangian to the Hamiltonian description of an open system in classical mechanics.


\begin{showData}
\section*{Data Availability}

Data sharing is not applicable to this article as no new data were created or analyzed in this study.
\end{showData}

\subsection*{Acknowledgements}

We thank Professor Leonid Polterovich for directing us to \cite{Daz} and \cite{Mar}, and for a conversation with Adam Yassine at a workshop at MSRI that guided us away from a fruitless direction.

\end{document}